\documentclass[reprint,aps,prd,nobibnotes,nofootinbib,onecolumn]{revtex4-2} 

\usepackage{amsmath}
\usepackage{amssymb}
\usepackage{graphicx}
\usepackage{fullpage}
\usepackage{hyperref}

\usepackage{color}
\definecolor{darkblue}{rgb}{0,0,0.5}

\begin{document}

\title{Flavor-Changing Non-Global Logarithms}%

\author{Andrew J.~Larkoski}%
\email{larkoski@aps.org}
\affiliation{American Physical Society, Hauppauge, New York 11788, USA}


\date{\today}%

\begin{abstract}
\noindent Non-global logarithms are low energy correlations between the substructure of a jet and the event in which it is immersed.  We study the leading non-global logarithms that arise from soft quark--anti-quark emission and calculate their coefficient as a series in the jet radius, $R$, in arbitrary processes.  We calculate the exact coefficient through quadratic order in $R$, and show that this truncation is within 5\% of the complete result for radii up to $R = 1$.  These quark flavor-dependent non-global logarithms are also responsible for the infrared unsafety of a na\"ive definition of jet flavor that is simply the net sum of quark flavors in the jet of interest.  We propose a small modification of this na\"ive jet flavor that we call {\it subtractive jet flavor} in which the problematic soft logarithms are explicitly subtracted.  We further demonstrate how our analytic results can be interfaced with automated numerical fixed-order codes to extract subtractive jet flavor cross sections.
\end{abstract}

\maketitle


\section{Introduction}

\noindent Jets, collimated streams of high energy particles produced in collider experiment, are manifestations of quantum chromodynamics (QCD).  Because of the property of asymptotic freedom, jets are proxies for the physics at short distances and through their observable particles, encode information of the initiating particle or particles.  With this in mind, significant effort has been devoted to development of observables that are sensitive to and can isolate properties of jets, and distinguish different classes of jets \cite{Salam:2010nqg,Abdesselam:2010pt,Altheimer:2012mn,Altheimer:2013yza,Larkoski:2017jix,Kogler:2018hem,Marzani:2019hun}.

Though identified as localized regions on the celestial sphere or in a detector experiment, jets do not exist in isolation.  Jets are simply part of a complete collision event, and can exhibit correlations between the particles that compose their substructure and the rest of the event.  The effect of some of these correlations are well-understood, like global momentum conservation, but there are other, more subtle sources of correlations that spoil the simple picture that a jet can always be mapped to a unique progenitor.  Because QCD is a weakly-coupled gauge theory, the dominant emissions in a jet are hard and collinear to the jet's core, or soft and emitted at wide angles.  In particular, these soft emissions can split further into additional soft and wide angle emissions, and one of those daughter emissions can land in the jet of interest and the other outside of it.  These leading soft correlations between the inside of a jet and the rest of the event in which it is immersed are referred to as non-global logarithms \cite{Dasgupta:2001sh}, as they exist exclusively because a jet's region is not the entire event.

The study of leading non-global logarithms has a rather long history, and are generated by strongly-ordered soft gluon emission.  Typically, one considers the scenario in which a measurement is made on the particles in the jet of interest, but is sufficiently inclusive over activity outside of the jet.  In this scenario, the evolution equation for non-global logarithms is non-linear \cite{Banfi:2002hw}, which greatly complicates a general, all-orders analysis.  Nevertheless, there have been several efforts to calculate these leading non-global logarithms to high orders in perturbation theory \cite{Schwartz:2014wha,Caron-Huot:2015bja,Larkoski:2015zka,Becher:2015hka,AngelesMartinez:2018cfz}, at subleading logarithmic accuracy \cite{Banfi:2021xzn,FerrarioRavasio:2023kyg,Becher:2023vrh}, or beyond the leading-color approximation \cite{Weigert:2003mm,Hatta:2013iba,Khelifa-Kerfa:2015mma}.

These studies of non-global logarithms have typically been restricted to an evolution that exclusively produces gluons, but this is not all that can happen.  At subleading accuracy, a soft gluon can be emitted that subsequently splits into a soft quark--anti-quark pair.  One of the quarks can then land in the jet, and the other outside.  However, because of angular momentum or fermion number conservation, there is no enhancement of the matrix element when a single quark becomes soft, and so non-global logarithms from soft quark emission are subdominant.  A detailed study of the non-global logarithms from soft quark--anti-quark emission is necessary for a precision understanding of a jet's structure and substructure.  In the context of hemisphere jets produced at a lepton collider, the leading quark--anti-quark non-global logarithms have long been known \cite{Kelley:2011ng,Hornig:2011iu}, but their generalization to jets in arbitrary processes has not been studied thoroughly.

Additionally, there has recently been renewed interest in robust definitions of jet flavor, theoretically well-defined techniques for classifying jets as flavors of quarks or as a gluon \cite{Banfi:2006hf,Caletti:2022hnc,Caletti:2022glq,Czakon:2022wam,Gauld:2022lem,Caola:2023wpj,Behring:2025ilo}.  Jet flavor is especially relevant for making predictions for heavy flavor production, to be able to match and compare a calculation to what is actually measured in experiment, e.g., a first study in Ref.~\cite{LHCb:2025tvf}.  In QCD perturbation theory, the simplest such definition of jet flavor that one could imagine is to sum together the individual flavors of quarks that land in the jet of interest.  This definition is well-defined at leading and next-to-leading order, but fails beginning at next-to-next-to-leading order, when there are two additional emissions in the event \cite{Banfi:2006hf}.  As they carry no quark flavor, gluon emissions never affect the flavor label of a jet, but at next-to-next-to-leading order, there exist soft quark emissions that can.  A soft quark--anti-quark pair can be emitted, with one of the quarks landing in the jet, and the other outside.  This soft quark changes the flavor sum of the particles in the jet, and further the soft quark can be arbitrarily low energy.  As such, this na\"ive flavor sum suffers from infrared divergences, and lacks the property of infrared and collinear (IRC) safety, necessary to ensure predictability within perturbation theory.

Both a detailed study of non-global logarithms and a robust jet flavor definition motivate understanding the distribution of soft quark emissions.  In this paper, we bring together these two directions, employing calculations of the non-global logarithms from soft quark--anti-quark emission for establishing an infrared safe jet flavor definition.  To the goal of understanding non-global logarithms, we approach the problem in generality, at leading power in the soft limit for the emission of a soft quark and anti-quark.  The most commonly-used jet algorithm at the Large Hadron Collider (LHC) is the anti-$k_T$ algorithm \cite{Cacciari:2008gp}, and in the soft limit, it reduces to a simple angular cut on the soft particles about the hard jet core.  We consider an arbitrary configuration of jets in the event, and calculate possible non-global logarithms emitted from any color dipole.

The probability that a soft quark lands in (or not) a jet depends on the size of that jet, as quantified by its radius $R$.  Calculating the complete jet radius dependence in the non-global logarithms is almost certainly not analytically tractable.  However, we are able to calculate order-by-order in powers of $R$, and explicitly present exact expressions through order $R^2$.  We compare these exact, though of truncated accuracy, results to numerical integration on the complete phase space and show that they agree to within 5\% for all $R \leq 1$, and to better than 1\% for $R\lesssim 0.5$.  Therefore, these simplified expressions are likely sufficiently accurate for precision jet studies in which non-global logarithms are present.

With our calculations of non-global logarithms in hand, we can then use them to define jet flavor.  Specifically, we propose a subtractive jet flavor algorithm in which we determine the net quark flavor of the jet, just as with na\"ive jet flavor, however, only those particles with sufficiently high energy contribute.  In the limit that the appropriate energy cut on the particles in the jet is lowered, the probabilities for any individual jet flavor are dominated by exactly these soft quark--anti-quark non-global logarithms.  To define an infrared safe jet flavor when completely inclusive over all particles in the jet, we then simply subtract off the divergences, subtracting off the flavor-dependent non-global logarithms.  While we are at present unable to make high-precision predictions with this flavor algorithm, we demonstrate the procedure through automated fixed-order event generation.

The outline of this paper is as follows.  In Sec.~\ref{sec:ngls}, we introduce flavor-dependent non-global logarithms, and our procedure for calculating them as a series in the jet radius $R$.  We present compact, analytic expressions for the non-global logarithms in a jet produced in $e^+e^-\to$ dijets, $pp\to V+$jet, and $pp\to $ dijets events, where $V$ is an electroweak boson.  In Sec.~\ref{sec:jetflav}, we revisit na\"ive jet flavor and present the subtractive jet flavor algorithm, given our results on non-global logarithms.  We demonstrate how this algorithm works at next-to-next-to-leading order, and show that our analytic formulae agree well with the non-global logarithms in $pp\to W^+ bu\bar u$ events, as generated in {\sc MadGraph} \cite{Alwall:2014hca}.  In Sec.~\ref{sec:allords}, we present some ideas about how the subtractive jet flavor can apply at all orders in perturbation theory, and correspondingly how it can be applied on a realistic jet consisting of an arbitrary number of particles.  We conclude in Sec.~\ref{sec:concs}, summarizing our results and looking forward to more applications. Appendices contain some additional results necessary for practical implementation of the subtractive jet flavor algorithm.

\section{Calculating Non-Global Logarithms as a series in the Jet Radius}\label{sec:ngls}

\begin{figure}[t!]
\begin{center}
\includegraphics[width=0.45\textwidth]{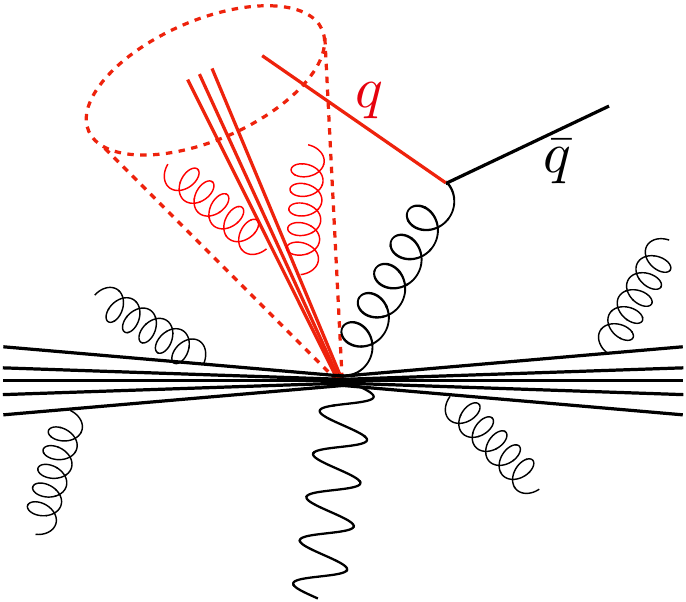}
\caption{\label{fig:ngl_config}
Illustration of the configuration we study in this paper in which non-global logarithms from soft quark emission appear.  All particles in the jet of interest are red, and the jet's region in the experiment is the dashed cone.  A soft gluon is emitted, which subsequently splits to a quark $q$ and anti-quark $\bar q$ pair.  The quark lands in the jet, while the anti-quark does not.
}
\end{center}
\end{figure}

Our goal will be to calculate the leading, flavor-dependent non-global logarithms as a series in the jet radius, $R$.  We will consider the case in which a soft $q\bar q$ pair is emitted from the hard jets in the event, one of the quarks lies in the jet, and the other lies outside the jet.  Further, we make a measurement of some infrared safe observable $\tau$ on the jet of interest, which constrains the energy of the soft quarks.  Note that the observable we measure on the jet for this situation does not need to be collinear safe, as all possible collinear divergences are regulated by the jet requirements.  This configuration we consider is illustrated in Fig.~\ref{fig:ngl_config}.

Given this set-up, the master formula for this calculation is
\begin{align}
\sigma^{\text{(2)}}_{q\bar q\text{,NGL}}(\tau) \equiv \int d\Pi_{q\bar q}\,|{\cal M}_{q\bar q}|^2\,\Theta\left(\hat \tau(\Pi_{q\bar q})-\tau\right)\Theta_\text{jet}\,.
\end{align}
Here, we have defined the cumulative cross section for measuring the observable $\tau$ on soft $q\bar q$ phase space $\Pi_{q\bar q}$ with squared matrix element $|{\cal M}_{q\bar q}|^2$.  To isolate the non-global configuration, we must enforce a jet clustering constraint $\Theta_\text{jet}$ that forces one of the soft quarks to lie in the jet of interest, and the other to be outside.  We will define or review each component in this master formula in the following.

Concretely, we will exclusively consider jet finding with the anti-$k_T$ algorithm \cite{Cacciari:2008gp}, so that our results can be directly applied to results from experiments at the LHC.  The pairwise clustering metric of anti-$k_T$ at a hadron collider is
\begin{align}
d_{ij} = \min\left[
p_{\perp,i}^{-2},p_{\perp,j}^{-2}
\right]\frac{(\phi_i-\phi_j)^2+(\eta_i-\eta_j)^2}{R^2}\,,
\end{align}
while the metric between a particle $i$ and the beams is
\begin{align}
d_{iB} = p_{\perp,i}^{-2}\,,
\end{align}
where $R$ is the jet radius, $p_\perp$ is the momentum transverse to the beam, $\phi$ is the azimuthal angle about the beam, and $\eta$ is the pseudorapidity along the beam.  To extract soft logarithms, we work to leading power in the energy or transverse momentum of the soft $q\bar q$ pair.  Jet clustering proceeds ordered from smallest pairwise metric $d_{ij}$ up, summing together particles' four momenta, until the beam distance is the smallest.  At this point, all particles that have been summed together form a jet with characteristic radius $R$.  Note that the anti-$k_T$ metric is minimized when the transverse momenta are large.  Two soft particles would have very large clustering metric $d_{ij}$, and so are not clustered.  Thus, the constraint that one quark lies in the jet, near a highly-energetic particle, and the other lies outside the jet reduces to exclusively an angular constraint.  Specifically, the in-jet/out-of-jet restriction is simply
\begin{align}
\Theta_{\text{jet}} = \Theta\left(R^2-(\eta_J-\eta_1)^2-(\phi_J-\phi_1)^2\right)\Theta\left((\eta_J-\eta_2)^2+(\phi_J-\phi_2)^2-R^2\right)\,,
\end{align}
where the soft quark and anti-quark are labeled as 1 and/or 2, and the jet of interest is $J$.

As mentioned above, because we restrict the soft quarks to not lie in the same jet, there are no collinear divergences.  Additionally, there is no strongly-ordered soft limit because quark number is conserved.  Therefore, the simplest possible observable that regulates the soft divergence is to measure the energy of one of the quark or anti-quark's transverse momentum.  As we have focused on a jet of interest $J$, we will ignore everything outside of it, and measure the transverse momentum of the soft particle that lies in the jet.  The observable $\tau$ is then 
\begin{align}
\hat \tau(\Pi_{q\bar q}) = p_{\perp,1}\,.
\end{align}
Our observable constraint is to simply require a lower bound on the transverse momentum in the jet, $p_{\perp,1}>p_{\perp,\text{cut}}$.\footnote{This constraint can be generalized to any IRC safe observable.  To the order that we work, all that is affected is the argument of the soft logarithm, modified from $p_{\perp,\text{cut}}$ to the general $\tau$.}

Phase space for two soft particles has no momentum conservation constraints, so takes the form
\begin{align}
\int d\Pi_{q\bar q} &= \int \frac{d^dp_1}{(2\pi)^d}\frac{d^dp_2}{(2\pi)^d}\,2\pi\delta(p_1^2)\,2\pi\delta(p_2^2) \\
&=\frac{1}{4(2\pi)^{6-4\epsilon}} \int d\eta_1\,p_{\perp,1}^{1-2\epsilon}dp_{\perp,1}\,d\phi_1\,\sin^{-2\epsilon}\phi_1 \,d\eta_2\,p_{\perp,2}^{1-2\epsilon}\,dp_{\perp,2}\,d\phi_2\,\sin^{-2\epsilon}\phi_2\nonumber\,,
\end{align}
in dimensional regularization, with $d=4-2\epsilon$.  In the $\overline{\text{MS}}$ scheme in which we work, we also will ignore factors of $(4\pi)^\epsilon$ in the following.  The squared matrix element for soft emission of a $q\bar q$ pair is 
\begin{align}
|{\cal M}_{q\bar q}|^2 \equiv (4\pi\mu^{2\epsilon}\alpha_s)^2 T_R\,\sigma_\text{hard} \sum_{i,j=1}^n {\cal I}_{ij}(p_1,p_2)\,.
\end{align}
Here, $\sigma_\text{hard}$ is the hard cross section describing the probability of the configuration of $n$ hard particles in the event, $T_R = 1/2$ is the normalization of the fundamental representation of SU(3) color, and $\mu$ is the renormalization scale.  The soft matrix element describing emission from dipole $ij$ is \cite{Catani:1999ss}
\begin{align}
{\cal I}_{ij}(p_1,p_2)&={\bf T}_i\cdot{\bf T}_j\,\frac{(n_i\cdot p_1)(n_j\cdot p_2)+(n_j\cdot p_1)(n_i\cdot p_2)-(n_i\cdot n_j)(p_1\cdot p_2)}{(p_1\cdot p_2)^2\left[n_i\cdot(p_1+p_2)\right]\left[n_j\cdot(p_1+p_2)\right]}\,.
\end{align}
$n_i = (1,\hat n_i)$ is a four-vector for hard particle $i$, where $\hat n_i$ is a unit three-vector in the direction of particle $i$.

Putting these pieces together, our master formula is
\begin{align}
\sigma^{\text{(2)}}_{q\bar q\text{,NGL}}(p_{\perp,\text{cut}})&=\left(
\frac{\alpha_s}{2\pi}
\right)^2\frac{\mu^{4\epsilon} T_R}{(2\pi)^{2}}\,\sigma_\text{hard}\sum_{i,j=1}^n{\bf T}_i\cdot{\bf T}_j\int d\eta_1\,p_{\perp,1}^{1-2\epsilon}dp_{\perp,1}\,d\phi_1\,\sin^{-2\epsilon}\phi_1 \,d\eta_2\,p_{\perp,2}^{1-2\epsilon}\,dp_{\perp,2}\,d\phi_2\,\sin^{-2\epsilon}\phi_2\nonumber\\
&\hspace{1cm}\times \frac{(n_i\cdot p_1)(n_j\cdot p_2)+(n_j\cdot p_1)(n_i\cdot p_2)-(n_i\cdot n_j)(p_1\cdot p_2)}{(p_1\cdot p_2)^2\left[n_i\cdot(p_1+p_2)\right]\left[n_j\cdot(p_1+p_2)\right]}\\
&\hspace{1cm}\times\Theta\left(R^2-(\eta_J-\eta_1)^2-(\phi_J-\phi_1)^2\right)\Theta\left((\eta_J-\eta_2)^2+(\phi_J-\phi_2)^2-R^2\right)\Theta\left(
p_{\perp,1}-p_{\perp,\text{cut}}
\right)\,.
\nonumber
\end{align}
The transverse momenta of both soft quarks can be integrated over.  Making the change of variables
\begin{align}
& p_{\perp,1}=p\,, &p_{\perp,2} = up\,,
\end{align}
the integral is homogeneous in $p$:
\begin{align}
\sigma^{\text{(2)}}_{q\bar q\text{,NGL}}(p_{\perp,\text{cut}})&=\left(
\frac{\alpha_s}{2\pi}
\right)^2\frac{\mu^{4\epsilon} T_R}{(2\pi)^{2}}\,\sigma_\text{hard}\sum_{i,j=1}^n{\bf T}_i\cdot{\bf T}_j\int d\eta_1\,\frac{dp}{p^{1+4\epsilon}}\,d\phi_1\,\sin^{-2\epsilon}\phi_1 \,d\eta_2\,u^{-2\epsilon}\,du\,d\phi_2\,\sin^{-2\epsilon}\phi_2\nonumber\\
&\hspace{1cm}\times \frac{(n_i\cdot \tilde p_1)(n_j\cdot \tilde p_2)+(n_j\cdot \tilde p_1)(n_i\cdot \tilde p_2)-(n_i\cdot n_j)(\tilde p_1\cdot \tilde p_2)}{(\tilde p_1\cdot \tilde p_2)^2\left[n_i\cdot(\tilde p_1+u\tilde p_2)\right]\left[n_j\cdot(\tilde p_1+u\tilde p_2)\right]}\\
&\hspace{1cm}\times\Theta\left(R^2-(\eta_J-\eta_1)^2-(\phi_J-\phi_1)^2\right)\Theta\left((\eta_J-\eta_2)^2+(\phi_J-\phi_2)^2-R^2\right)\Theta\left(
p-p_{\perp,\text{cut}}
\right)\nonumber\\
&=\left(
\frac{\alpha_s}{2\pi}
\right)^2\left(
\frac{\mu^2}{p_{\perp,\text{cut}}^2}
\right)^{2\epsilon}\frac{T_R}{(2\pi)^{2}}\,\frac{\sigma_\text{hard}}{4\epsilon}\sum_{i,j=1}^n{\bf T}_i\cdot{\bf T}_j\int d\eta_1\,d\phi_1\,\sin^{-2\epsilon}\phi_1 \,d\eta_2\,u^{-2\epsilon}\,du\,d\phi_2\,\sin^{-2\epsilon}\phi_2\nonumber\\
&\hspace{1cm}\times \frac{(n_i\cdot \tilde p_1)(n_j\cdot \tilde p_2)+(n_j\cdot \tilde p_1)(n_i\cdot \tilde p_2)-(n_i\cdot n_j)(\tilde p_1\cdot \tilde p_2)}{(\tilde p_1\cdot \tilde p_2)^2\left[n_i\cdot(\tilde p_1+u\tilde p_2)\right]\left[n_j\cdot(\tilde p_1+u\tilde p_2)\right]}\nonumber\\
&\hspace{1cm}\times\Theta\left(R^2-(\eta_J-\eta_1)^2-(\phi_J-\phi_1)^2\right)\Theta\left((\eta_J-\eta_2)^2+(\phi_J-\phi_2)^2-R^2\right)\,.
\nonumber
\end{align}
In this final expression, we have introduced the dimensionless vectors $\tilde p$ for the soft quarks, which is just their four-vector stripped of the overall transverse momentum factor:
\begin{align}
\tilde p = (\cosh\eta,\cos\phi,\sin\phi,\sinh\eta)\,.
\end{align}
Further, note that the integrals that remain are completely finite because of the jet restriction, so we can set $\epsilon=0$ within the integral.  We than have 
\begin{align}
\sigma^{\text{(2)}}_{q\bar q\text{,NGL}}(p_{\perp,\text{cut}})
&=\left(
\frac{\alpha_s}{2\pi}
\right)^2\left(
\frac{\mu^2}{p_{\perp,\text{cut}}^2}
\right)^{2\epsilon}\frac{T_R}{(2\pi)^{2}}\,\frac{\sigma_\text{hard}}{4\epsilon}\sum_{i,j=1}^n{\bf T}_i\cdot{\bf T}_j\int d\eta_1\,d\phi_1 \,d\eta_2\,d\phi_2\,du\\
&\hspace{2cm}\times \frac{(n_i\cdot \tilde p_1)(n_j\cdot \tilde p_2)+(n_j\cdot \tilde p_1)(n_i\cdot \tilde p_2)-(n_i\cdot n_j)(\tilde p_1\cdot \tilde p_2)}{(\tilde p_1\cdot \tilde p_2)^2\left[n_i\cdot(\tilde p_1+u\tilde p_2)\right]\left[n_j\cdot(\tilde p_1+u\tilde p_2)\right]}\nonumber\\
&\hspace{2cm}\times\Theta\left(R^2-(\eta_J-\eta_1)^2-(\phi_J-\phi_1)^2\right)\Theta\left((\eta_J-\eta_2)^2+(\phi_J-\phi_2)^2-R^2\right)\,.
\nonumber
\end{align}

The integral over the transverse momentum ratio $u$ can also be done.  The result depends on the structure of the dipole, where
\begin{align}
&\int du\, \frac{(n_i\cdot \tilde p_1)(n_j\cdot \tilde p_2)+(n_j\cdot \tilde p_1)(n_i\cdot \tilde p_2)-(n_i\cdot n_j)(\tilde p_1\cdot \tilde p_2)}{(\tilde p_1\cdot \tilde p_2)^2\left[n_i\cdot(\tilde p_1+u\tilde p_2)\right]\left[n_j\cdot(\tilde p_1+u\tilde p_2)\right]}\\
&\hspace{4cm}=\left\{
\begin{array}{cc}
\frac{2{\bf T}_i^2}{(\tilde p_1\cdot \tilde p_2)^2}\,,  &\text{if }i=j\\
\\
{\bf T}_i\cdot {\bf T}_j\,\frac{(n_i\cdot \tilde p_1)(n_j\cdot \tilde p_2)+(n_j\cdot \tilde p_1)(n_i\cdot \tilde p_2)-(n_i\cdot n_j)(\tilde p_1\cdot \tilde p_2)}{(\tilde p_1\cdot \tilde p_2)^2\left[
(n_i\cdot \tilde p_1)(n_j\cdot \tilde p_2)-(n_j\cdot \tilde p_1)(n_i\cdot \tilde p_2)
\right]} \log\frac{(n_i\cdot \tilde p_1)(n_j\cdot \tilde p_2)}{(n_j\cdot \tilde p_1)(n_i\cdot \tilde p_2)}  \,, &\text{if }i\neq j
\end{array}
\right.\nonumber
\end{align}
By color conservation, we note that
\begin{align}
\sum_{i=1}^n {\bf T}_i^2 = -2\sum_{1\leq i<j\leq n}{\bf T}_i\cdot {\bf T}_j\,,
\end{align}
so these results can be combined into a single term where
\begin{align}
\sigma^{\text{(2)}}_{q\bar q\text{,NGL}}(p_{\perp,\text{cut}})
&=\left(
\frac{\alpha_s}{2\pi}
\right)^2\left(
\frac{\mu^2}{p_{\perp,\text{cut}}^2}
\right)^{2\epsilon}\frac{T_R}{2\pi^{2}}\,\frac{\sigma_\text{hard}}{4\epsilon}\sum_{1\leq i<j\leq n}{\bf T}_i\cdot{\bf T}_j\int d\eta_1\,d\phi_1 \,d\eta_2\,d\phi_2\\
&\hspace{0.5cm}\times \frac{1}{(\tilde p_1\cdot \tilde p_2)^2}\left[
\frac{(n_i\cdot \tilde p_1)(n_j\cdot \tilde p_2)+(n_j\cdot \tilde p_1)(n_i\cdot \tilde p_2)-(n_i\cdot n_j)(\tilde p_1\cdot \tilde p_2)}{(n_i\cdot \tilde p_1)(n_j\cdot \tilde p_2)-(n_j\cdot \tilde p_1)(n_i\cdot \tilde p_2)} \log\frac{(n_i\cdot \tilde p_1)(n_j\cdot \tilde p_2)}{(n_j\cdot \tilde p_1)(n_i\cdot \tilde p_2)}-2
\right]\nonumber\\
&\hspace{0.5cm}\times\Theta\left(R^2-(\eta_J-\eta_1)^2-(\phi_J-\phi_1)^2\right)\Theta\left((\eta_J-\eta_2)^2+(\phi_J-\phi_2)^2-R^2\right)\,.
\nonumber
\end{align}

Before working to evaluate the angular integrals, there are a few simplifications that can be exploited.  First, note that the matrix element is invariant to rescalings of the unit four-vectors $n_i$ and $n_j$.  Also, the integration measure is invariant to translations in both pseudorapidity (boosts along the beam) and azimuth (rotations about the beam).  Therefore, for evaluation of this soft function, we can boost all hard jets to the frame in which the jet of interest lies at $\eta_J=\phi_J = 0$.  The cross section then simplifies to
\begin{align}
\sigma^{\text{(2)}}_{q\bar q\text{,NGL}}(p_{\perp,\text{cut}})
&=\left(
\frac{\alpha_s}{2\pi}
\right)^2\left(
\frac{\mu^2}{p_{\perp,\text{cut}}^2}
\right)^{2\epsilon}\frac{T_R}{2\pi^{2}}\,\frac{\sigma_\text{hard}}{4\epsilon}\sum_{1\leq i<j\leq n}{\bf T}_i\cdot{\bf T}_j\int d\eta_1\,d\phi_1 \,d\eta_2\,d\phi_2\\
&\hspace{0.5cm}\times \frac{1}{(\tilde p_1\cdot \tilde p_2)^2}\left[
\frac{(n_i\cdot \tilde p_1)(n_j\cdot \tilde p_2)+(n_j\cdot \tilde p_1)(n_i\cdot \tilde p_2)-(n_i\cdot n_j)(\tilde p_1\cdot \tilde p_2)}{(n_i\cdot \tilde p_1)(n_j\cdot \tilde p_2)-(n_j\cdot \tilde p_1)(n_i\cdot \tilde p_2)} \log\frac{(n_i\cdot \tilde p_1)(n_j\cdot \tilde p_2)}{(n_j\cdot \tilde p_1)(n_i\cdot \tilde p_2)}-2
\right]\nonumber\\
&\hspace{0.5cm}\times\Theta\left(R^2-\eta_1^2-\phi_1^2\right)\Theta\left(\eta_2^2+\phi_2^2-R^2\right)\,.
\nonumber
\end{align}
For explicit calculations that follow, we also note that the dot products in the integrand are:
\begin{align}
&\tilde p_1\cdot \tilde p_2= \cosh(\eta_1-\eta_2)-\cos(\phi_1-\phi_2)\,,
&n_i\cdot \tilde p\propto \cosh(\eta-\eta_i)-\cos(\phi-\phi_i)\,.
\end{align}
Also, in what follows, it will be helpful to have a compact notation for the integrand, so we repurpose the soft matrix element, where
\begin{align}
{\cal I}_{ij}(\tilde p_1,\tilde p_2) \equiv \frac{{\bf T}_i\cdot{\bf T}_j}{(\tilde p_1\cdot \tilde p_2)^2}\left[
\frac{(n_i\cdot \tilde p_1)(n_j\cdot \tilde p_2)+(n_j\cdot \tilde p_1)(n_i\cdot \tilde p_2)-(n_i\cdot n_j)(\tilde p_1\cdot \tilde p_2)}{(n_i\cdot \tilde p_1)(n_j\cdot \tilde p_2)-(n_j\cdot \tilde p_1)(n_i\cdot \tilde p_2)} \log\frac{(n_i\cdot \tilde p_1)(n_j\cdot \tilde p_2)}{(n_j\cdot \tilde p_1)(n_i\cdot \tilde p_2)}-2
\right]\,.
\end{align}

\subsection{Correlations between the jet of interest and another jet}

Now, there are two possibilities for the evaluation of this integral.  First, one of jets $i$ or $j$ could be the jet of interest, $J$, or both $i$ and $j$ could be other jets in the event.  We will consider the case when $i=J$, say, first, and in the following section study the (more complicated) case of when $i,j\neq J$.  Our goal will be to calculate the cross section through second order in the jet radius, ${\cal O}(R^2)$.  We will do this through two steps.  First, we will expand the matrix element ${\cal I}_{ij}(\tilde p_1,\tilde p_2)$ in the collinear limit, where we assume that both $\tilde p_1$ and $\tilde p_2$ are an angle of order $R$ from the jet of interest.  Second, while quark 1 is required to be close to the jet of interest, quark 2 is not, so we construct the contribution when quark 2 is a large angle from the jet of interest.  Then, the total contribution is simply their sum.

We start with the expansion in the collinear limit.  To isolate this region, we rescale all angular coordinates by the jet radius $R$, where
\begin{align}
&\eta_1 \to R\eta_1\,, &\phi_1\to R\phi_1\,,\\
&\eta_2 \to R\eta_2\,, &\phi_2\to R\phi_2\,.
\end{align}
Then, with this rescaling, we expand the matrix element through order-$R^{-2}$, where 
\begin{align}
{\cal I}_{JA}(R\eta_1,R\phi_1,R\eta_2,R\phi_2) = R^{-4}{\cal I}^{\text{(coll,0)}}_{JA}(\eta_1,\phi_1,\eta_2,\phi_2)+R^{-2}{\cal I}^{\text{(coll,2)}}_{JA}(\eta_1,\phi_1,\eta_2,\phi_2)+{\cal O}(R^0)\,.
\end{align}
Here, $A$ denotes any other jet in the event.  Note that the integration measure scales like $R^4$, so expansion through this order ensures terms through an overall cross section scaling of $R^2$ are captured.  The leading collinear matrix element is rather simple, where
\begin{align}
{\cal I}^{\text{(coll,0)}}_{JA}(\eta_1,\phi_1,\eta_2,\phi_2) = \frac{8\,{\bf T}_J\cdot {\bf T}_A}{\left(
(\eta_1-\eta_2)^2+(\phi_1-\phi_2)^2
\right)^2}\left[
\frac{\eta_1\eta_2+\phi_1\phi_2}{\eta_1^2-\eta_2^2+\phi_1^2-\phi_2^2}\log\frac{\eta_1^2+\phi_1^2}{\eta_2^2+\phi_2^2}-1
\right]\,,
\end{align}
and in particular, note that it is independent of the direction of the other jet $A$.  The subleading collinear matrix element, ${\cal I}^{\text{(coll,2)}}_{JA}(\eta_1,\phi_1,\eta_2,\phi_2)$, by contrast, is not so simple, so we will not write it down here explicitly.  However, the integral over them is rather simple, and we find
\begin{align}
&\int d\eta_1\,d\phi_1\,d\eta_2\,d\phi_2\left[
{\cal I}^{\text{(coll,0)}}_{JA}(\eta_1,\phi_1,\eta_2,\phi_2)+{\cal I}^{\text{(coll,2)}}_{JA}(\eta_1,\phi_1,\eta_2,\phi_2)
\right]\Theta\left(R^2-\eta_1^2-\phi_1^2\right)\Theta\left(\eta_2^2+\phi_2^2-R^2\right)\\
&\hspace{4cm}={\bf T}_J\cdot {\bf T}_A\left[
\frac{4}{3}\pi^2-\frac{8}{9}\pi^4+4R^2+\frac{\pi^2}{3}R^2-2\pi^2R^2\frac{\cos\phi_A}{\cosh\eta_A-\cos\phi_A}+{\cal O}(R^4)
\right]\,.\nonumber
\end{align}

Now, we turn to the contribution from the region when quark 2 is an order-1 angle from the jet of interest, assuming $R\ll 1$.  This region can be isolated by rescaling only quark 1's direction:
\begin{align}
&\eta_1\to R\eta_1\,,
&\phi_1\to R\phi_1\,,
\end{align}
in the matrix element.  However, we also need to eliminate the double counting from the collinear region already calculated, so we subtract the collinear matrix elements with the same scaling.  That is, the matrix element for a wide-angle quark 2 is
\begin{align}
{\cal I}^{\text{(wide)}}_{JA}(\eta_1,\phi_1,\eta_2,\phi_2)&\equiv{\cal I}_{JA}(R\eta_1,R\phi_1,\eta_2,\phi_2)-{\cal I}^{\text{(coll,0)}}_{JA}(R\eta_1,R\phi_1,\eta_2,\phi_2)-{\cal I}^{\text{(coll,2)}}_{JA}(R\eta_1,R\phi_1,\eta_2,\phi_2)\\
&={\cal I}^{\text{(wide,0)}}_{JA}(\eta_2,\phi_2)+{\cal O}(R^2)\,.\nonumber
\end{align}
Note that here we need to expand to order-$R^0$ because the integration measure with this scaling is itself order-$R^2$.  Further, because this matrix element scales like $R^0$, it is necessarily independent of the (small) coordinates $\eta_1,\phi_1$.  This subtracted matrix element is also very simple, where
\begin{align}
{\cal I}^{\text{(wide,0)}}_{JA}(\eta_2,\phi_2) = {\bf T}_J\cdot{\bf T}_A\left[
\frac{8}{\left(\eta_2^2+\phi_2^2\right)^2}-\frac{4}{3}\frac{\eta_2^2-\phi_2^2}{\left(\eta_2^2+\phi_2^2\right)^2}-\frac{2}{\left(\cosh\eta_2-\cos\phi_2\right)^2}
\right]\,.
\end{align}
Its integral over all of particle 2 phase space is
\begin{align}
\pi R^2\int d\eta_2\,d\phi_2\, {\cal I}^{\text{(wide,0)}}_{JA}(\eta_2,\phi_2)={\bf T}_J\cdot {\bf T}_A\left[
-4R^2-\frac{4}{3}\pi^2 R^2
\right]\,.
\end{align}

Now, putting these pieces together, the complete contribution to the non-global cross section from correlations between the jet of interest $J$ and another jet $A$ is
\begin{align}
&\int d\eta_1\,d\phi_1\,d\eta_2\,d\phi_2\, {\cal I}_{JA}(\eta_1,\phi_1,\eta_2,\phi_2)\,\Theta(R^2-\eta_1^2-\phi_1^2)\,\Theta(\eta_2^2+\phi_2^2-R^2) \\
&\hspace{4cm}= {\bf T}_J\cdot {\bf T}_A\left[
\frac{4}{3}\pi^2-\frac{8}{9}\pi^4-\pi^2R^2\,\frac{\cosh\eta_A+\cos\phi_A}{\cosh\eta_A-\cos\phi_A}+{\cal O}(R^4)
\right]\,.\nonumber
\end{align}
This can be cast in an invariant way, in terms of dot products of direction vectors of the jets.  We equivalently have
\begin{align}
&\int d\eta_1\,d\phi_1\,d\eta_2\,d\phi_2\, {\cal I}_{JA}(\eta_1,\phi_1,\eta_2,\phi_2)\,\Theta(R^2-\eta_1^2-\phi_1^2)\,\Theta(\eta_2^2+\phi_2^2-R^2) \\
&\hspace{5cm}= {\bf T}_J\cdot {\bf T}_A\left[
\frac{4}{3}\pi^2-\frac{8}{9}\pi^4-\pi^2R^2\,\frac{\bar n_J\cdot n_A}{n_J\cdot n_A}+{\cal O}(R^4)
\right]\,.\nonumber
\end{align}
where $\bar n_J = (1,-\hat n_J)$, so that $n_J\cdot \bar n_J = 2$.

\subsection{Correlations between two other jets}

We now turn our attention to evaluation of the angular integrals from the dipole consisting of jets $A$ and $B$, which are both distinct from the jet of interest, $J$.  As earlier, we consider the collinear expansion, and then the region when quark 2 is emitted at wide angle.  Performing the rescalings established earlier, the collinear matrix element is
\begin{align}
{\cal I}_{AB}(R\eta_1,R\phi_1,R\eta_2,R\phi_2) = R^{-2}{\cal I}^{\text{(coll,0)}}_{AB}(\eta_1,\phi_1,\eta_2,\phi_2)+{\cal O}(R^0)\,.
\end{align}
Note that now, there is no contribution at order-$R^{-4}$ because other jets in the event are assumed to lie an order-1 angle from one another.  Additionally, while this expansion is simple enough to take, the resulting matrix element is extremely unwieldy, and we have not been able to find a compact expression for it.  As such, we will not write it down here.  Nevertheless, its integral over phase space does simplify nicely, where
\begin{align}\label{eq:collnj-nj}
&\int d\eta_1\,d\phi_1\,d\eta_2\,d\phi_2\,  {\cal I}^{\text{(coll,0)}}_{AB}(\eta_1,\phi_1,\eta_2,\phi_2)\,\Theta\left(R^2-\eta_1^2-\phi_1^2\right)\Theta\left(\eta_2^2+\phi_2^2-R^2\right)\\
&\hspace{2cm}={\bf T}_A\cdot {\bf T}_B\left[
\frac{4}{3}\pi^2\,\frac{n_A\cdot n_B}{(n_A\cdot n_J)(n_B\cdot n_J)}\left(
R^2\log\frac{R^2}{4\pi^2}-R^2
\right)\right.\nonumber\\
&\hspace{3cm}\left.+
\frac{16}{3}\pi^2 R^2\,\frac{\Re\left[\sinh^2\left(\frac{\eta_A-i\phi_A}{2}
\right)\sinh^2\left(\frac{\eta_B -i\phi_B}{2}
\right)\sinh^2\left(\frac{\eta_A-\eta_B +i(\phi_A-\phi_B)}{2}
\right)\right]}{(\cosh\eta_A-\cos\phi_A)^2(\cosh\eta_B-\cos\phi_B)^2}+{\cal O}(R^4)\right]\,.\nonumber
\end{align}
On the second line, the result has been expressed in a coordinate-independent way, while on the third line, we have not found a coordinate-independent way to express it, and so explicit $\eta,\phi$ dependence is presented.  Here, $\Re$ denotes the real part.

For the wide-angle contribution, constructing this matrix element is rather easy.  Because jets $A$ and $B$ are a wide-angle from jet $J$ themselves, to leading power in the small angle limit for quark 1, we simply set its direction to be exactly along the jet of interest, $\tilde p_1 = n_J$.  To eliminate double counting with the collinear region, we then subtract the collinear matrix element ${\cal I}^{\text{(coll,0)}}_{AB}(\tilde p_1,\tilde p_2)$, with the same identification for quark 1.  That is, the wide angle matrix element for this dipole is
\begin{align}
{\cal I}^{\text{(wide,0)}}_{AB}(\tilde p_1,\tilde p_2) = {\cal I}_{AB}(\tilde p_1=n_J,\tilde p_2)-{\cal I}^{\text{(coll,0)}}_{AB}(\tilde p_1=n_J,\tilde p_2)\,.
\end{align}
The collinear subtraction matrix element can be integrated on phase space and expressed as an explicit function of all three relevant jet angular coordinates, $J$, $A$ and $B$, using standard symbolic integration software (i.e., {\sc Mathematica}).  However, this completely general expression is again unwieldy, so will not be written here explicitly.  

By contrast, we were unable to evaluate the integral over ${\cal I}_{AB}(\tilde p_1=n_J,\tilde p_2)$ with the native routines in {\sc Mathematica}.  However, studying the integral more closely, it can be evaluated.  Let's write down the relevant integral, where
\begin{align}
&\pi R^2\int d\eta_2\, d\phi_2\, {\cal I}_{AB}(\tilde p_1=n_J,\tilde p_2) \\
&= \pi R^2\int d\eta_2\, d\phi_2\,\frac{{\bf T}_A\cdot{\bf T}_B}{(n_J\cdot \tilde p_2)^2}\left[
\frac{(n_A\cdot n_J)(n_B\cdot \tilde p_2)+(n_B\cdot n_J)(n_A\cdot \tilde p_2)-(n_A\cdot n_B)(n_J\cdot \tilde p_2)}{(n_A\cdot n_J)(n_B\cdot \tilde p_2)-(n_B\cdot n_J)(n_A\cdot \tilde p_2)} \log\frac{(n_A\cdot n_J)(n_B\cdot \tilde p_2)}{(n_B\cdot n_J)(n_A\cdot \tilde p_2)}-2
\right]\nonumber\,.
\end{align}
The overall factor of $\pi R^2$ is the area of the jet of interest, where quark 1 lies.  The integrand, as written, is technically divergent (hence the need for the collinear subtraction), but this is a bit of a distraction.  The divergence can be regulated by making any cutoff on integration variables in the region around $\eta_2,\phi_2\to 0$.  Performing the same cutoff prescription on the collinear subtraction then ensures that the result is independent of the cutoff. What we will do here is to integrate over all $\phi_2\in[0,2\pi)$, but restrict $|\eta_2|>R$.  Note that this expression is already at order-$R^2$, and so any contribution from the missing region is further suppressed by $R$, and beyond our approximation.

\begin{figure}[t!]
\begin{center}
\includegraphics[width=0.55\textwidth]{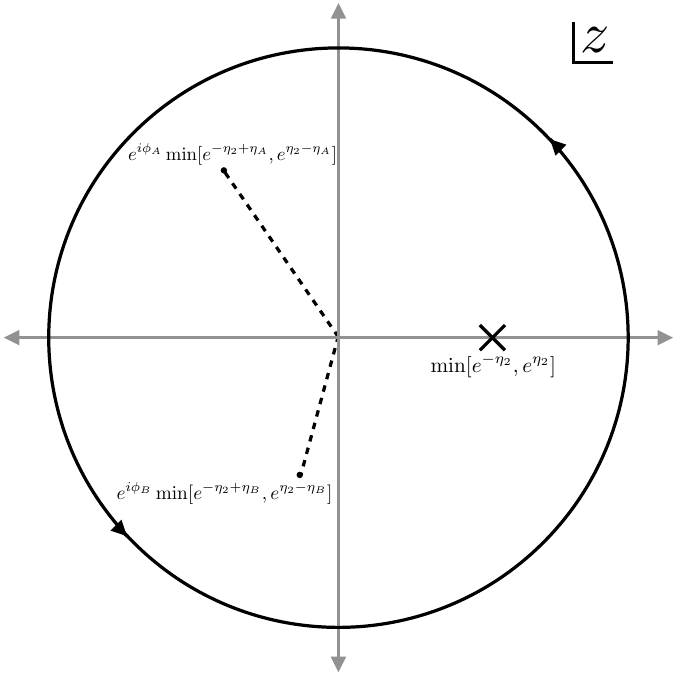}
\caption{\label{fig:complex}
Illustration of the non-analytic structure within the contour (solid black) about the unit circle in the complex $z=e^{i\phi_2}$ plane.  There is a pole located at $z_*=\min[e^{-\eta_2},e^{\eta_2}]$ (cross) and logarithmic branch cuts (dashed) along the line segments $z_\text{$A$,cut}\in\left[0,e^{i\phi_A}\min[e^{\eta_2-\eta_A},e^{-\eta_2+\eta_A}]\right]$ and $z_\text{$B$,cut}\in\left[0,e^{i\phi_B}\min[e^{\eta_2-\eta_B},e^{-\eta_2+\eta_B}]\right]$.
}
\end{center}
\end{figure}

Then, with this regularization, the integral is
\begin{align}
&\pi R^2\int d\eta_2\, d\phi_2\, {\cal I}_{AB}(\tilde p_1=n_J,\tilde p_2) \\
&\hspace{1cm}= \pi R^2\int_{|\eta_2|>R} d\eta_2\int_0^{2\pi} d\phi_2\,\frac{{\bf T}_A\cdot{\bf T}_B}{(n_J\cdot \tilde p_2)^2}\nonumber\\
&\hspace{2cm}\times\left[
\frac{(n_A\cdot n_J)(n_B\cdot \tilde p_2)+(n_B\cdot n_J)(n_A\cdot \tilde p_2)-(n_A\cdot n_B)(n_J\cdot \tilde p_2)}{(n_A\cdot n_J)(n_B\cdot \tilde p_2)-(n_B\cdot n_J)(n_A\cdot \tilde p_2)} \log\frac{(n_A\cdot n_J)(n_B\cdot \tilde p_2)}{(n_B\cdot n_J)(n_A\cdot \tilde p_2)}-2
\right]\nonumber\,.
\end{align}
We can then change variables from $\phi_2$ to complex $z$, where
\begin{align}
z = e^{i\phi_2}\,,
\end{align}
and perform the contour integral about the unit circle.  The non-analytic structure inside the unit circle is rather simple in this case.  First, there is only a single pole located at $z_* = \min[e^{\eta_2},e^{-\eta_2}]$, which comes from the overall factor $1/(n_J\cdot \tilde p_2)^2$.  Next, there are two branch cuts from the logarithms $\log(n_A\cdot \tilde p_2)$ and $\log(n_B\cdot \tilde p_2)$.  These branch cuts live on the line segments $z_\text{$A$,cut}\in\left[0,e^{i\phi_A}\min[e^{\eta_2-\eta_A},e^{-\eta_2+\eta_A}]\right]$ and $z_\text{$B$,cut}\in\left[0,e^{i\phi_B}\min[e^{\eta_2-\eta_B},e^{-\eta_2+\eta_B}]\right]$, respectively.  The non-analytic structure and contour on the $z$ plane we consider are displayed in Fig.~\ref{fig:complex}.  Evaluating the corresponding residues or integrals along the branch cuts is straightforward, and can be done analytically.  The integral over $\eta_2$ can then be correspondingly evaluated; however, again, the final expression we find is horribly unwieldy and we have not been able to simplify it, so we do not report it here.

Instead, we will present results for two relevant, simple cases for the orientation of the jets $A$ and $B$.  First, we consider $A$ and $B$ to be the colliding beams, which live at pseudorapidity $\pm \infty$.  The matrix element for this configuration is very simple, where
\begin{align}
{\cal I}_{AB}^\text{(beams)}(\tilde p_1=n_J,\tilde p_2) =\frac{2\,{\bf T}_A\cdot {\bf T}_B}{(\cosh\eta_2-\cos\phi_2)^2}\left[\frac{\cos\phi_2}{\sinh\eta_2} \,\eta_2 -1\right]\,.
\end{align}
The integral of the wide-angle matrix element for the beams correlation is
\begin{align}
&\pi R^2 \int d\eta_2\, d\phi_2\, \left[{\cal I}_{AB}^\text{(beams)}(\tilde p_1=n_J,\tilde p_2)-{\cal I}^{\text{(beams,coll,0)}}_{AB}(\tilde p_1=n_J,\tilde p_2)\right]\\
&\hspace{6cm}={\bf T}_A\cdot {\bf T}_B\left[-\frac{64}{9}\pi^2 R^2
+\frac{8}{3}\pi^2R^2\log(4\pi^2)
\right]\,.\nonumber
\end{align}

The second configuration we consider is between a beam $A$, at, say, pseudorapidity $+\infty$, and a second jet $J_2$, back-to-back in the transverse plane of the jet of interest, $J$.  This second jet then lives at $\phi_{J_2} = \pi$, with arbitrary pseudorapidity $\eta_{J_2}$. The corresponding matrix element is
\begin{align}
{\cal I}_{A{J_2}}^\text{(beam-jet 2)}(\tilde p_1=n_J,\tilde p_2) &=\frac{{\bf T}_A\cdot {\bf T}_{J_2}}{(\cosh\eta_2-\cos\phi_2)^2}\\
&\hspace{-3.5cm}\times\left[\frac{\cosh(\eta_2-\eta_{J_2})+\cos\phi_2+(\cosh\eta_{J_2}+1)e^{-\eta_2}-e^{-\eta_{J_2}}(\cosh\eta_2-\cos\phi_2)}{\cosh(\eta_2-\eta_{J_2})+\cos\phi_2-(\cosh\eta_{J_2}+1)e^{-\eta_2}} \log\frac{\cosh(\eta_2-\eta_{J_2})+\cos\phi_2}{(\cosh\eta_{J_2}+1)e^{-\eta_2}} -2\right]\,.\nonumber
\end{align}
Its collinear-subtracted integral over phase space is
\begin{align}
&\pi R^2 \int d\eta_2\, d\phi_2\, \left[{\cal I}_{A{J_2}}^\text{(beam-jet 2)}(\tilde p_1=n_J,\tilde p_2)-{\cal I}^{\text{(beam-jet 2,coll,0)}}_{A{J_2}}(\tilde p_1=n_J,\tilde p_2)\right]\\
&\hspace{4cm}={\bf T}_A\cdot {\bf T}_{J_2}\,e^{-\eta_{J_2}}\text{sech}^2\frac{\eta_{J_2}}{2}\left[-\frac{16}{9}\pi^2 R^2
-\frac{4}{3}\pi^2 R^2\log\left(\frac{1+e^{\eta_{J_2}}}{2}\right)+\frac{2}{3}\pi^2R^2\log\pi^2
\right]\,.\nonumber
\end{align}
As discussed above, this integral can be evaluated by contour integration.  

With these results, we can then combine them to present the beam-beam and beam-jet contributions.  Combining with Eq.~\ref{eq:collnj-nj}, we have
\begin{align}\label{eq:beambeamcorrfin}
&\int d\eta_1\,d\phi_1\,d\eta_2\,d\phi_2\,{\cal I}_{AB}^\text{(beams)}(\tilde p_1,\tilde p_2)\,\Theta(R^2-\eta_1^2-\phi_1^2)\,\Theta(\eta_2^2+\phi_2^2-R^2) \\
&\hspace{4cm}={\bf T}_A\cdot {\bf T}_B\left[
-\frac{76}{9}\pi^2 R^2+\frac{8}{3}\pi^2 R^2\log R^2+{\cal O}(R^4)
\right]\,,
\nonumber
\end{align}
for the correlations between the two beams.  For the correlation between a beam and the jet back-to-back from the jet of interest, we find
\begin{align}\label{eq:jetbeamcorr}
&\int d\eta_1\,d\phi_1\,d\eta_2\,d\phi_2\,{\cal I}_{AJ_2}^\text{(beam-jet)}(\tilde p_1,\tilde p_2)\,\Theta(R^2-\eta_1^2-\phi_1^2)\,\Theta(\eta_2^2+\phi_2^2-R^2) \\
&\hspace{3cm}={\bf T}_A\cdot {\bf T}_{J_2}\,e^{-\eta_{J_2}}\text{sech}^2\frac{\eta_{J_2}}{2}\left[
-\frac{19}{9}\pi^2 R^2-\frac{4}{3}\pi^2 R^2\log\left(
1+e^{\eta_{J_2}}
\right)+\frac{2}{3}\pi^2R^2\log R^2+{\cal O}(R^4)
\right]\,.\nonumber
\end{align}
These two configurations enable evaluation of the flavor-dependent non-global logarithms to one- and two-jet processes.  Note also that Eq.~\ref{eq:jetbeamcorr} agrees with Eq.~\ref{eq:beambeamcorrfin} when the jet $J_2$ becomes the other beam, $\eta_{J_2}\to-\infty$.

\subsection{Non-Global Logarithms in Three Different Events}

Given these general results, we will apply them to evaluating the flavor-dependent non-global logarithms in three different classes of events.  We again note that the bare contribution to the cross section from soft quark emission split across the jet boundary is
\begin{align}
\sigma^{\text{(2)}}_{q\bar q\text{,NGL}}(p_{\perp,\text{cut}})
&=\left(
\frac{\alpha_s}{2\pi}
\right)^2\left(
\frac{\mu^2}{p_{\perp,\text{cut}}^2}
\right)^{2\epsilon}\frac{T_R}{2\pi^{2}}\,\frac{\sigma_\text{hard}}{4\epsilon}\sum_{1\leq i<j\leq n}{\bf T}_i\cdot{\bf T}_j\int d\eta_1\,d\phi_1 \,d\eta_2\,d\phi_2\\
&\hspace{0.5cm}\times \frac{1}{(\tilde p_1\cdot \tilde p_2)^2}\left[
\frac{(n_i\cdot \tilde p_1)(n_j\cdot \tilde p_2)+(n_j\cdot \tilde p_1)(n_i\cdot \tilde p_2)-(n_i\cdot n_j)(\tilde p_1\cdot \tilde p_2)}{(n_i\cdot \tilde p_1)(n_j\cdot \tilde p_2)-(n_j\cdot \tilde p_1)(n_i\cdot \tilde p_2)} \log\frac{(n_i\cdot \tilde p_1)(n_j\cdot \tilde p_2)}{(n_j\cdot \tilde p_1)(n_i\cdot \tilde p_2)}-2
\right]\nonumber\\
&\hspace{0.5cm}\times\Theta\left(R^2-\eta_1^2-\phi_1^2\right)\Theta\left(\eta_2^2+\phi_2^2-R^2\right)\,.
\nonumber
\end{align}
This exhibits an explicit soft divergence in dimensional regularization, but this will be subtracted when combined with the appropriate virtual contributions.  We will thus focus on the renormalized soft cross section
\begin{align}
\sigma^{\text{(2)}}_{q\bar q\text{,NGL}}(p_{\perp,\text{cut}})
&=\sigma_\text{hard}\left(
\frac{\alpha_s}{2\pi}
\right)^2\frac{T_R}{2\pi^{2}}\log\frac{\mu}{p_{\perp,\text{cut}}}\sum_{1\leq i<j\leq n}{\bf T}_i\cdot{\bf T}_j\int d\eta_1\,d\phi_1 \,d\eta_2\,d\phi_2\\
&\hspace{0.5cm}\times \frac{1}{(\tilde p_1\cdot \tilde p_2)^2}\left[
\frac{(n_i\cdot \tilde p_1)(n_j\cdot \tilde p_2)+(n_j\cdot \tilde p_1)(n_i\cdot \tilde p_2)-(n_i\cdot n_j)(\tilde p_1\cdot \tilde p_2)}{(n_i\cdot \tilde p_1)(n_j\cdot \tilde p_2)-(n_j\cdot \tilde p_1)(n_i\cdot \tilde p_2)} \log\frac{(n_i\cdot \tilde p_1)(n_j\cdot \tilde p_2)}{(n_j\cdot \tilde p_1)(n_i\cdot \tilde p_2)}-2
\right]\nonumber\\
&\hspace{0.5cm}\times\Theta\left(R^2-\eta_1^2-\phi_1^2\right)\Theta\left(\eta_2^2+\phi_2^2-R^2\right)\,,
\nonumber
\end{align}
where $\mu$ is the renormalization scale, and is also the scale at which the strong coupling $\alpha_s$ is evaluated.  Additionally, we note that, as written so far, this cross section only contains the contribution from a single quark flavor, $q$.  This point will be important in the latter half of this paper.  However, in general, one would typically consider the cross section inclusive over all $n_f$ light quark flavors, and so this cross section should be multiplied by a factor of $n_f$ in those cases.

\subsubsection{$e^+e^-\to jj$}

While our general calculation was formulated for jets produced in $pp$ collisions, our results also apply for dijet production in $e^+e^-$ collisions.  Non-global logarithms from soft quarks split across the event hemisphere have long been calculated \cite{Kelley:2011ng,Hornig:2011iu}, so this application is more of a check of our calculation here.  In this simple case, there is only a single dipole between the jet of interest and the other jet in the event, and further, those jets are back-to-back in the center of mass frame.  The non-global logarithms for this configuration are then
\begin{align}
\sigma^{\text{(2)}}_{q\bar q\text{,NGL}}(p_{\perp,\text{cut}})
&=\sigma_{e^+e^-\to jj}^{(0)}\left(
\frac{\alpha_s}{2\pi}
\right)^2T_R\,{\bf T}_J\cdot {\bf T}_{J_2}\,\log\frac{\mu}{p_{\perp,\text{cut}}}\left[
\frac{2}{3}-\frac{4}{9}\pi^2+{\cal O}(R^4)
\right]
\,,
\end{align}
where $\sigma_{e^+e^-\to jj}^{(0)}$ is the leading order cross section for jet production in $e^+e^-$ collisions.  This agrees with known results, and further, there are no order-$R^2$ corrections to the result in the collinear limit.  Again, this is well-known and not surprising, as those earlier calculations used an event's hemispheres as jets, and so the jet radius was as large as possible, $R = \pi/2$.

\subsubsection{$pp\to V+j$}

Next, we consider these non-global logarithms in $pp\to V+j$ events, where $V$ is an electroweak boson and $j$ is the recoiling jet.  As it is the simplest non-trivial example, we will study this event configuration throughout the rest of this paper, and in particular will test the analytic expression through order-$R^2$ for these non-global logarithms to the complete numerical integral expression, as well as in event simulation.

This event has three hard color lines, the two colliding beams $A$ and $B$ and the jet of interest in the final state, $J$.  Summing together the various contributions from the previous sections, we have
\begin{align}
\sigma^\text{(2)}_{q\bar q\text{,NGL}}(p_{\perp,\text{cut}})
&=\sigma_{pp\to V+j}^{(0)}\left(
\frac{\alpha_s}{2\pi}
\right)^2T_R\,\log\frac{\mu}{p_{\perp,\text{cut}}}\left[
{\bf T}_A\cdot {\bf T}_B\left(-\frac{38}{9} R^2
+\frac{4}{3}R^2\log R^2\right)\right.\\
&\hspace{6cm}\left.+\,{\bf T}_J\cdot \left({\bf T}_A+{\bf T}_B\right)\left(
\frac{2}{3}-\frac{4}{9}\pi^2-\frac{R^2}{2}
\right)
+{\cal O}(R^4)
\right]
\,,\nonumber
\end{align}
where $\sigma_{pp\to V+j}^{(0)}$ is the leading-order cross section for this process.  Using color conservation, this can be written exclusively in terms of quadratic Casimirs of SU(3) color.  We note that 
\begin{align}
{\bf T}_A+{\bf T}_B= - {\bf T}_J\,,
\end{align}
and 
\begin{align}
{\bf T}_A\cdot {\bf T}_B = \frac{({\bf T}_A+ {\bf T}_B)^2-{\bf T}_A^2-{\bf T}_B^2}{2} = \frac{{\bf T}_J^2-{\bf T}_A^2-{\bf T}_B^2}{2}\,.
\end{align}
The non-global logarithms can then be expressed as 
\begin{align}\label{eq:vjnglquad}
\sigma^\text{(2)}_{q\bar q\text{,NGL}}(p_{\perp,\text{cut}})
&=\sigma_{pp\to V+j}^{(0)}\left(
\frac{\alpha_s}{2\pi}
\right)^2T_R\,\log\frac{\mu}{p_{\perp,\text{cut}}}\left[
\left({\bf T}_A^2+ {\bf T}_B^2\right)\left(\frac{19}{9} R^2
-\frac{2}{3}R^2\log R^2\right)\right.\\
&\hspace{5.5cm}\left.+\,{\bf T}_J^2\left(
-\frac{2}{3}+\frac{4}{9}\pi^2-\frac{29}{18} R^2
+\frac{2}{3}R^2\log R^2
\right)
+{\cal O}(R^4)
\right]
\,.\nonumber
\end{align}

\subsubsection{$pp\to jj$}

From the presented general results, we are also able to construct the non-global logarithms for $pp\to jj$ events.  As discussed earlier, we have boosted the jet of interest along the beam line to live at 0 pseudorapidity, and rotated about the beam to be at 0 azimuth.  Therefore, at leading order by transverse momentum conservation, the second jet necessarily lies at azimthal angle $\pi$, but may have non-zero pseudorapidity due to a residual boost along the beams.  This configuration then indicates the necessity of convolving the leading-order cross section for $pp\to jj$ events as a function of jet pseudorapidities with the soft matrix element, with inclusive jet selection.  Such a convolution is beyond what we will consider here, but we can express the non-global logarithms differential in the second jet's pseudorapidity.  

With this configuration and jet selection, we can use the result of Eq.~\ref{eq:jetbeamcorr} to calculate the complete expression for non-global logarithms to this configuration.  We have
\begin{align}
\sigma^\text{(2)}_{q\bar q\text{,NGL}}(p_{\perp,\text{cut}})
&=\sigma_{pp\to jj}^{(0)}(\eta_{J_2})\left(
\frac{\alpha_s}{2\pi}
\right)^2T_R\,\log\frac{\mu}{p_{\perp,\text{cut}}}\left[
{\bf T}_A\cdot {\bf T}_B\left(-\frac{38}{9} R^2
+\frac{4}{3}R^2\log R^2\right)\right.\\
&\hspace{1cm}+\,{\bf T}_J\cdot \left({\bf T}_A+{\bf T}_B\right)\left(
\frac{2}{3}-\frac{4}{9}\pi^2-\frac{R^2}{2}
\right)+\,{\bf T}_J\cdot {\bf T}_{J_2}\left(
\frac{2}{3}-\frac{4}{9}\pi^2-\frac{R^2}{2}\frac{\cosh\eta_{J_2}-1}{\cosh\eta_{J_2}+1}
\right)\nonumber\\
&\hspace{1cm}+\,{\bf T}_A\cdot {\bf T}_{J_2}\,e^{-\eta_{J_2}}\text{sech}^2\frac{\eta_{J_2}}{2}\left(
-\frac{19}{18}R^2-\frac{2}{3}R^2\log\left(
1+e^{\eta_{J_2}}
\right)+\frac{1}{3}R^2\log R^2
\right)\nonumber\\
&\hspace{1cm}\left.+\,{\bf T}_B\cdot {\bf T}_{J_2}\,e^{\eta_{J_2}}\text{sech}^2\frac{\eta_{J_2}}{2}\left(
-\frac{19}{18} R^2-\frac{2}{3} R^2\log\left(
1+e^{-\eta_{J_2}}
\right)+\frac{1}{3}R^2\log R^2
\right)
+{\cal O}(R^4)
\right]
\,,\nonumber
\end{align}
where $\sigma_{pp\to jj}^{(0)}(\eta_{J_2})$ is the leading-order cross section to this process as a function of the pseudorapidity of the other jet.  Color conservation can simplify this expression, but knowledge of the full color matrices are needed to completely evaluate this expression.

\subsection{Numerical Comparisons}

Given these results as an expansion in the jet radius, we would like to determine the extent to which this truncation is a good approximation.  Na\"ively, we would expect that the approximation begins to break down for $R\sim 1$ as neglected terms of order $R^4$ would presumably be important by then.  However, in other contexts, it has been demonstrated that a narrow-jet or small-radius approximation remains accurate up to $R\sim 1$, see, e.g., Refs.~\cite{Aversa:1990uv,Jager:2004jh,Dasgupta:2007wa,Hornig:2011tg,Dasgupta:2012hg,Dasgupta:2013ihk}, so the same numerical coincidences may exist here.  First, as already mentioned, the collinear limit of the non-global logarithms is actually the complete result for $e^+e^-\to jj$ events, and so the truncation at order $R^2$ is also exactly correct.  For a non-trivial test, we must then go beyond to more complicated events.

We therefore consider $pp\to V+j$ events and will compare our truncation to the numerical integration of the soft $q\bar q$ matrix element with complete phase space restrictions.  Specifically, we will compare the truncation and numerical integration of the non-global logarithms expressed in the form of Eq.~\ref{eq:vjnglquad}, with quadratic Casimirs.  In this form, as the cross section must be positive, each term must also be positive.  We will separate this complete expression into two contributions, depending on its color structure.  First, we consider the contribution proportional to the jet's Casimir, which can be expressed as the integral over the matrix elements:
\begin{align}\label{eq:vjjet}
&\int d\eta_1\,d\phi_1\,d\eta_2\,d\phi_2\,\left[-{\cal I}_{JA}(\tilde p_1,\tilde p_2)-{\cal I}_{JB}(\tilde p_1,\tilde p_2)+\frac{1}{2}\,{\cal I}_{AB}(\tilde p_1,\tilde p_2)\right]\,\Theta(R^2-\eta_1^2-\phi_1^2)\,\Theta(\eta_2^2+\phi_2^2-R^2) \\
&\hspace{4cm}\supset{\bf T}_J^2\left[
-\frac{4}{3}\pi^2+\frac{8}{9}\pi^4-\frac{29}{9}\pi^2 R^2+\frac{4}{3}\pi^2 R^2\log R^2+{\cal O}(R^4)
\right]\,.
\nonumber
\end{align}
Second, we consider the contribution proportional to the sum of the beams' Casimirs, where
\begin{align}\label{eq:vjbeam}
&\int d\eta_1\,d\phi_1\,d\eta_2\,d\phi_2\,\left[-\frac{1}{2}\,{\cal I}_{AB}(\tilde p_1,\tilde p_2)\right]\,\Theta(R^2-\eta_1^2-\phi_1^2)\,\Theta(\eta_2^2+\phi_2^2-R^2) \\
&\hspace{4cm}\supset\left({\bf T}_A^2+{\bf T}_B^2\right)\left[
\frac{38}{9}\pi^2 R^2-\frac{4}{3}\pi^2 R^2\log R^2+{\cal O}(R^4)
\right]\,.
\nonumber
\end{align}

\begin{figure}[t!]
\begin{center}
\includegraphics[width=0.45\textwidth]{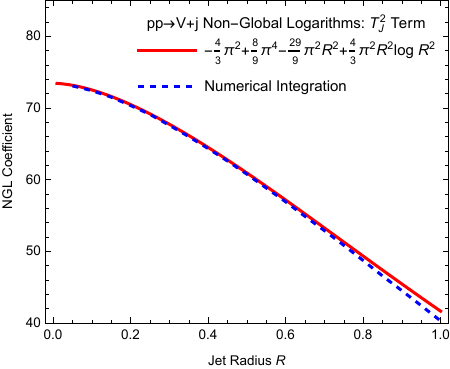}\hspace{1cm}
\includegraphics[width=0.45\textwidth]{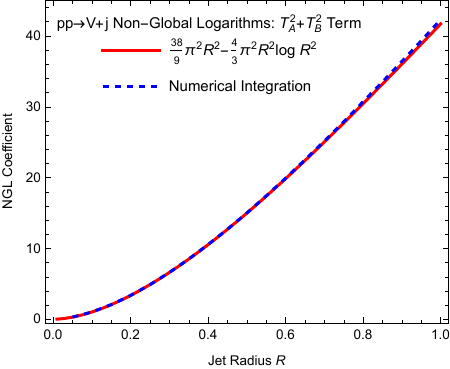}
\caption{\label{fig:nglnumerical}
Comparisons of the order-$R^2$ truncation of the flavor-dependent non-global logarithms in $pp\to V+j$ events versus the complete contribution as calculated through numerical integration, as a function of the jet radius $R$.  Left: The term proportional to ${\bf T}_J^2$, the Casimir of the jet.  Right: The term proportional to ${\bf T}_A^2+{\bf T}_B^2$, the sum of Casimirs of the colliding beams.
}
\end{center}
\end{figure}

To perform the numerical integrals expressed in Eqs.~\ref{eq:vjjet} and \ref{eq:vjbeam}, we use the implementation of the {\sc Vegas} algorithm \cite{Lepage:1977sw,Lepage:1980dq} in {\sc Cuba} 4.2 \cite{Hahn:2004fe}, for a range of jet radii $R\in[0,1]$.  We then directly compare the results of that numerical integration to the expansions through order-$R^2$, as written above.  The results of this comparison are plotted in Fig.~\ref{fig:nglnumerical}.  In general, the order-$R^2$ truncation agrees extremely well with the full numerical integral, even out to $R = 1$.  In particular, the approximation lies within 5\% of the complete result for all $R \lesssim 1$, and well within 1\% for $R \lesssim 0.5$.

A rather amusing feature of these results is that the contribution proportional to the jet's Casimir decreases with radius $R$, while the contribution proportional to the beams' Casimirs increases with $R$, such that by about $R = 1$, they are equal.  Thus, if one just used the collinear limit $R \to 0$ to approximate these non-global logarithms for jets with $R \sim 1$, which are often used in experimental analyses of high-mass jets, e.g., Refs.~\cite{CMS:2025eyd,CMS:2025bxo,ATLAS:2025bpp,ATLAS:2025rfm,ATLAS:2025woc}, one would have a poor approximation of their impact on a phenomenological study.

\section{Resurrecting Na\"ive Jet Flavor}\label{sec:jetflav}

Robust definitions of the partonic flavor of a jet has a long history \cite{Banfi:2006hf}, and has seen significant interest recently \cite{Caletti:2022hnc,Caletti:2022glq,Gauld:2022lem,Czakon:2022wam,Caola:2023wpj}.  For jet flavor to be useful and a systematically-improvable quantity it must be in particular infrared safe, insensitive to arbitrarily soft emissions.  Collinear safety is not strictly required, as one can define a jet's flavor through the leading partons produced through fragmentation, and potential divergences can be resummed through a fragmentation function.  In various ways, modern jet flavor algorithms ensure infrared safety of jet flavor, and this robustness is especially important for predicting and measuring jets with heavy flavor.

Our purpose here will not be to review or compare jet flavor algorithms.  Instead, given the study of flavor-dependent non-global logarithms, we will revisit the issues with the most na\"ive jet flavor prescription and present a simple solution.  Further, we are not arguing that this simple solution is an optimal algorithm, but rather is a baseline or benchmark for any possible jet flavor definition.  As we will demonstrate, if nothing else, this definition of jet flavor is simple to calculate in perturbation theory.

In QCD perturbation theory, a jet consists of a collection of quarks and gluons.  The most natural definition for the ``flavor'' of the jet is simply as the sum of the flavors of all particles in the jet.  Specifically, a jet $J$'s flavor vector, $\vec f_J$, is
\begin{align}
\vec f_J \equiv \sum_{i\in J} \vec f_i\,,
\end{align}
where $\vec f_i$ is the flavor of parton $i$ in the jet.  Specifically, $\vec f_i$ is an $n_f$ dimensional vector, where $n_f$ is the number of relevant quark flavors.  For a quark (anti-quark) of flavor $i$, $\vec f_i$ is a one-hot vector with 1 ($-1$) in the $i$th entry.  If $i=g$ is a gluon, then $\vec f_g = \vec 0$, the zero vector.  A jet's flavor vector then counts the net flavor of each type of quark in the jet.

There are various modifications that one can make to this general definition (like counting quark flavor mod 2 to account for neutral hadron oscillations \cite{mod2flav,Behring:2025ilo}), but we will work with this definition for simplicity.  This flavor definition is exactly what you would call the jet flavor at leading-order, when the jet has only a single particle in it.  At next-to-leading order, the only possible infrared divergence arises from soft gluon emission that subsequently lands in the jet.  However, as gluons carry no flavor, this does not and cannot change the flavor from its leading-order assignment, virtual and real divergences exactly cancel, and this flavor sum is infrared safe through this order.

Starting at next-to-next-to-leading order, the picture changes, however.  Still, soft gluon emissions are irrelevant, but now, a soft gluon can split into a soft quark--anti-quark pair.  The quark, say, can land in the jet region, while the anti-quark can be outside.  As the simple jet flavor sum has no energy weighting, the quark--anti-quark pair can be arbitrarily soft and affect the jet's flavor.  As such, this na\"ive jet flavor sum ceases to be infrared safe starting at next-to-next-to-leading order \cite{Banfi:2006hf}.  This configuration of emissions is exactly as illustrated in Fig.~\ref{fig:ngl_config}, in which flavor-dependent non-global logarithms appear.

\subsection{A Subtractive Jet Flavor Algorithm}

This observation, that flavor-dependent non-global logarithms are exactly what spoils the infrared safety of na\"ive jet flavor, then presents an intriguing solution.  If the infrared divergences that plague na\"ive jet flavor can be isolated, then we can explicitly subtract them given the known flavor-dependent non-global logarithms.  This motivates the following algorithm as an infrared safe modification to nai\"ve jet flavor, which we refer to as {\it subtractive jet flavor}.  The algorithm is as follows:
\begin{enumerate}

\item Cluster jets in your event with any infrared and collinear safe jet algorithm.  For applications to physics at the LHC, this is almost always the anti-$k_T$ algorithm \cite{Cacciari:2008gp}.  With applications to the LHC in mind, in the following, we will exclusively reference quantities appropriate for a hadron collider, but these can be simply modified to a different collider as appropriate.

\item On your jet of interest, compute the jet flavor, as a function of a minimum transverse momentum cut, $p_{\perp,\text{cut}}$.  That is, partons only contribute to the net jet flavor if their transverse momentum is greater than $p_{\perp,\text{cut}}$:
\begin{align}
\vec f_J(p_{\perp,\text{cut}}) \equiv \sum_{\substack{i\in J\\ p_{\perp,i} > p_{\perp,\text{cut}}}} \vec f_i\,.
\end{align}
 
\item On your event ensemble, now calculate the cross section for jet flavor given this transverse momentum cut, $\sigma_{\vec f_J}\left(p_{\perp,\text{cut}}\right)$.  In this cross section, as $p_{\perp,\text{cut}}\to 0$, there will be logarithms that would spoil smooth extrapolation to $p_{\perp,\text{cut}}= 0$.  These logarithms are exactly the flavor-dependent non-global logarithms, $\sigma_{\vec f_J,\text{NGL}}(p_{\perp,\text{cut}})$.  Define the cross section for subtractive jet flavor (SJF) by their difference:
\begin{align}
\sigma_\text{SJF}(\vec f_J) \equiv \lim_{p_{\perp,\text{cut}}\to 0}\left[ \sigma_{\vec f_J}\left(p_{\perp,\text{cut}}\right) - \sigma_{\vec f_J,\text{NGL}}(p_{\perp,\text{cut}})\right]\,.
\end{align}
This cross section is finite in the limit and is therefore infrared safe.

\end{enumerate}

In step 2., note that individual partonic energy cuts are not collinear safe, but this is a bit of a distraction, and not a fundamental limitation.  In particular, because we take the $p_{\perp,\text{cut}}\to 0$ limit, collinear divergences vanish at the end, but may arise in intermediate steps.  To eliminate them, one can introduce an appropriate fragmentation function that absorbs potential collinear divergences.  We construct the necessary fragmentation functions at one-loop order in App.~\ref{app:fragfunc}.  However, we also note that the fragmentation functions trivialize for $p_{\perp,\text{cut}}\to 0$.  Additionally, we note that such intermediate collinear divergences first arise at next-to-next-to-next-to-leading order (NNNLO) in perturbation theory, where there is a gluon emission, in addition to the soft quark--anti-quark pair.  At this order, there is a single collinear divergence in this final state configuration when the gluon is collinear to the soft quark that lies in the jet.  Therefore, the one-loop fragmentation functions can be used to subtract the collinear divergences that arise when imposing a finite $p_{\perp,\text{cut}}$.  At higher orders, fragmentation functions at higher loops will be necessary to subtract collinear divergences.

We additionally note that subtractive jet flavor is very much a fixed-order flavor definition.  Specifically, by definition, possible large logarithms are explicitly subtracted.  This subtraction means that there are no large logarithms that may spoil perturbative convergence and require all-orders resummation.  Therefore, potential collinear divergences in intermediate steps in the calculation of the subtractive jet flavor cross section can be dealt with on a case-by-case basis, order-by-order in perturbation theory that one calculates.  As mentioned above, this is first an issue at NNNLO, which is beyond current state-of-the-art precision calculations relevant for jet physics.

In a practical implementation of this procedure, one would typically evaluate the jet flavor as a function of $p_{\perp,\text{cut}}$, for a range of small but finite and non-zero values, in a numerical fixed-order code.  We present a first analysis of this procedure in Sec.~\ref{sec:flavmad} using {\sc MadGraph}, but it is worth commenting on numerical limitations here.  At next-to-next-to-leading order, when this flavor definition is first non-trivial, there are only single soft logarithms that need to be subtracted to extract the finite limiting value as $p_{\perp,\text{cut}}\to 0$.  This is therefore much simpler, and correspondingly much more numerically stable, than other cases in which fixed-order codes are used to extract finite residual contributions.  For example, high-precision fixed-order codes have long been used to extract two-loop constants in factorization theorems for infrared and collinear safe observables, e.g., for thrust in Ref.~\cite{Becher:2008cf}.  However, in the case of the thrust observable for example, even the next-to-leading order cross section as the value of thrust goes to 0 is a fourth-order polynomial in the logarithm of the thrust, and so a substantially more delicate cancellation is necessary to extract the finite contribution.  In Sec.~\ref{sec:flavmad}, we demonstrate that using the default settings in {\sc MadGraph} can produce reasonable extractions of subtractive jet flavor cross sections, even with no attempt at numerical optimization.

\subsubsection{Construction through NNLO}

Before applying this jet flavor definition in a concrete application, let's observe how it works at the first non-trivial order.  Note that the subtractive flavor algorithm is identical to na\"ive jet flavor at leading and next-to-leading order, while differences first arise at next-to-next-to-leading order.  We will therefore simply enumerate all possible parton configurations that can be present in the jet of interest at next-to-next-to-leading order and observe how the subtractive flavor algorithm works.

At next-to-next-to-leading order, a jet can consist of 1, 2, or 3 particles.  If the jet has 1 particle, its flavor is unambiguous and is simply the flavor of that one particle.  If the jet has 3 particles, then there are no possible soft final state emissions outside of the jet, so soft emissions contribute net 0 flavor to the jet.  The non-trivial case then, is if the jet consists of 2 particles, and there is another emission that lies outside the jet.  The case when the jet consists of two gluons is unambiguous: the jet is gluon flavor.  The interesting flavor configurations are: $qg$, $q\bar q$, and $qq'$, where $q$ and $q'$ are quarks with $q'\neq \bar q$.

With these two-particle configurations, we can enumerate the possible contributions to jet flavor cross sections, according to the transverse momenta of the relevant quarks.  We have:
\begin{align}
\sigma_g^{(2)}(p_{\perp,\text{cut}}) &\supset \sigma^{(2)}_{(qg)}\left(p_{\perp,q}<p_{\perp,\text{cut}}\right) + \sigma^{(2)}_{(q\bar q)}(p_{\perp,q},p_{\perp,\bar q}>p_{\perp,\text{cut}})\,,\\
\sigma_q^{(2)}(p_{\perp,\text{cut}}) &\supset\sigma^{(2)}_{(qg)}\left(p_{\perp,q}>p_{\perp,\text{cut}}\right)+\sigma^{(2)}_{(q\bar q)}(p_{\perp,\bar q}<p_{\perp,\text{cut}}<p_{\perp,q})+\sigma^{(2)}_{(q q')}(p_{\perp,q'}<p_{\perp,\text{cut}}<p_{\perp,q})\,,\\
\sigma_{(qq')}^{(2)}(p_{\perp,\text{cut}}) &\supset \sigma^{(2)}_{(q q')}(p_{\perp,q},p_{\perp,q'}>p_{\perp,\text{cut}})\,.
\end{align}
In these expressions, the superscript $(2)$ denotes the order in perturbation theory; i.e., next-to-next-to-leading order, the subscripts of the cross sections at farthest left denote the flavor bin, and the subscripts of the cross sections at right denote the jet's parton content.  For example, $\sigma^{(2)}_{(qg)}\left(p_{\perp,q}<p_{\perp,\text{cut}}\right)$ is the cross section for a jet with a quark $q$ and a gluon $g$ and contributes to the gluon jet flavor bin because the quark's transverse momentum is sufficiently small.  In these expressions, we also suppress implicit sums over unresolved quarks for compactness.

Now, according to the subtractive flavor algorithm, we take the $p_{\perp,\text{cut}}\to 0$ limit of these expressions.  We then have
\begin{align}
\lim_{p_{\perp,\text{cut}}\to 0} \sigma_g^{(2)}(p_{\perp,\text{cut}})&\supset -\sigma_{(gq)\bar q,\text{NGL}}^{(2)}(p_{\perp,\text{cut}})+\left[
\sigma^{(2)}_{(q\bar q)}(p_{\perp,q},p_{\perp,\bar q}>p_{\perp,\text{cut}}) - \sigma_{(q\bar q)q,\text{NGL}}^{(2)}(p_{\perp,\text{cut}})
\right]\\
&\hspace{9cm}+\sigma_{(q\bar q)q,\text{NGL}}^{(2)}(p_{\perp,\text{cut}})\,,\nonumber\\
\lim_{p_{\perp,\text{cut}}\to 0} \sigma_q^{(2)}(p_{\perp,\text{cut}}) &\supset \left[
\sigma^{(2)}_{(qg)}\left(p_{\perp,q}>p_{\perp,\text{cut}}\right) - \sigma_{(gq)\bar q,\text{NGL}}^{(2)}(p_{\perp,\text{cut}})
\right]+\sigma_{(gq)\bar q,\text{NGL}}^{(2)}(p_{\perp,\text{cut}})\\
&\hspace{6cm} - \sigma_{(q\bar q)q,\text{NGL}}^{(2)}(p_{\perp,\text{cut}})- \sigma_{(qq')\bar q',\text{NGL}}^{(2)}(p_{\perp,\text{cut}})\,,
\nonumber\\
\lim_{p_{\perp,\text{cut}}\to 0} \sigma_{(qq')}^{(2)}(p_{\perp,\text{cut}}) &\supset \left[
\sigma^{(2)}_{(q q')}(p_{\perp,q},p_{\perp,q'}>p_{\perp,\text{cut}}) - \sigma_{(qq')\bar q',\text{NGL}}^{(2)}(p_{\perp,\text{cut}})
\right]+\sigma_{(qq')\bar q',\text{NGL}}^{(2)}(p_{\perp,\text{cut}})\,.
\end{align}
Here, we have explicitly isolated the non-global logarithm contributions.  In particular, the flavor subscripts denote the flavor structure in the jet and event; for example, $(gq)\bar q$ means that the jet consists of a gluon $g$ and a quark $q$, and outside the jet is the anti-quark $\bar q$.  In this particular form, the quantities in the square brackets are finite as $p_{\perp,\text{cut}}\to 0$.  Further, note that all of the explicit non-global logarithmic contributions, those outside the square brackets, explicitly cancel in an inclusive sum over all individual jet flavor cross sections.  Of course, this must be true, as a sufficiently inclusive cross section has no large logarithms.

Finally, to construct the subtractive jet flavor algorithm, we subtract all those bare non-global logarithmic contributions.  We then have
\begin{align}
\sigma_\text{SJF}(g) &\supset \lim_{p_{\perp,\text{cut}}\to 0} \left[
\sigma^{(2)}_{(q\bar q)}(p_{\perp,q},p_{\perp,\bar q}>p_{\perp,\text{cut}}) - \sigma_{(q\bar q)q,\text{NGL}}^{(2)}(p_{\perp,\text{cut}})
\right]\,,\\
\sigma_\text{SJF}(q) &\supset\lim_{p_{\perp,\text{cut}}\to 0} \left[
\sigma^{(2)}_{(qg)}\left(p_{\perp,q}>p_{\perp,\text{cut}}\right) - \sigma_{(gq)\bar q,\text{NGL}}^{(2)}(p_{\perp,\text{cut}})
\right]\,,
\\
\sigma_\text{SJF}(qq') &\supset\lim_{p_{\perp,\text{cut}}\to 0} \left[
\sigma^{(2)}_{(q q')}(p_{\perp,q},p_{\perp,q'}>p_{\perp,\text{cut}}) - \sigma_{(qq')\bar q',\text{NGL}}^{(2)}(p_{\perp,\text{cut}})
\right]\,.
\end{align}
As constructed, all of these contributions to individual jet flavor bins are finite.  Again, these configurations are responsible for the infrared unsafety of na\"ive jet flavor, and all other contributions to the appropriate jet flavor can be identified by simply performing the flavor sum as discussed earlier.

One important point to note is that even after taking the $p_{\perp,\text{cut}}\to0$ limit, the infrared poles of the real emissions are still (implicitly) present in the subtractive flavor cross sections.  However, the transverse momentum cut and limit ensures that the real poles are associated with the correct flavor prescriptions such that they will cancel exactly with the corresponding virtual divergences.  Additionally, there will be finite low-scale matrix element contributions that remain from the $p_{\perp,\text{cut}}\to 0$ limit that are implicit.  Because these finite contributions arise from real, soft emission, they can be calculated once and for all and applied to a general process, just as we calculated the non-global logarithms in a general process.  In a practical implementation of this algorithm, one has to take finite values of $p_{\perp,\text{cut}}$ and then extrapolate to 0.  As such, these finite low-scale constants will always be included, and one can safely just subtract off the logarithmic divergence accordingly.

\subsection{Subtractive Jet Flavor at Fixed-Order}\label{sec:flavmad}

To fully validate and test the subtractive flavor algorithm requires complete predictions through next-to-next-to-leading order.  While there are now several studies with these modern flavor algorithms at this order \cite{Gauld:2020deh,Czakon:2020coa,Gehrmann-DeRidder:2023gdl,Mazzitelli:2024ura,Biello:2024pgo}, we will restrict an analysis here to a sufficiently simple flavor configuration that illustrates the subtleties with non-global logarithms, but does so only at next-to-leading order.  Specifically, we will consider the process $pp\to W^++j$ at the 13 TeV LHC, and focus on the jet flavor $f_j=(bu)$, a composite flavor in which there is both a sufficiently hard bottom and up quark.  This jet flavor configuration is chosen so that there is a single non-global logarithmic contribution, from the final state $W^+b u\bar u$.  In particular, note that the final state $W^+b\bar b u$ is not possible at tree level which simplifies our analysis here.

To study this process, we generate fixed-order events in {\sc MadGraph} v3.6.4 \cite{Alwall:2014hca} with default settings, except for what we discuss below.  By default, the bottom quark is too massive to have a distribution in the proton, so the process $pp\to W^+bu$ is only non-zero with a non-trivial CKM matrix.  For extraction of the non-global logarithms and comparison to our calculations, we need the leading-order cross section for the process $pp\to W^+b$.  We will further impose a transverse momentum cut on the $W^+$ boson of $p_{\perp,W} > 1000$ GeV, which is inclusive over the hadronic activity, and so is especially simple to impose at higher orders.  In {\sc MadGraph}, we find the leading-order cross section to be
\begin{align}\label{eq:lowb}
\sigma^{(0)}_{pp\to W^+b}(p_{\perp,W} > 1000\text{ GeV}) = 1.14\times 10^{-3}\text{ fb}\,.
\end{align}
Numerical uncertainty on the cross section is beyond the reported digits, and here, we will not worry about scale or theoretical uncertainties.

\subsubsection{Leading Order Analysis}

\begin{figure}[t!]
\begin{center}
\includegraphics[width=0.45\textwidth]{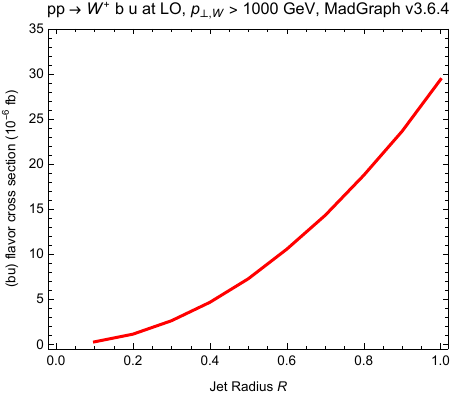}
\caption{\label{fig:loflavxsec}
Plot of the leading order cross section for $(bu)$ jet flavor from $pp\to W^+bu$ events with $p_{\perp,W} > 1000$ GeV, as generated in {\sc MadGraph} v3.6.4.  The jet contributes to the cross section for this flavor if the $b$ and $u$ are clustered within the jet radius $R$.
}
\end{center}
\end{figure}

We can calculate the leading-order cross section for $(bu)$ flavor production in this process, as a function of the jet radius $R$.  With only two particles, any $k_T$-style algorithm \cite{Ellis:1993tq,Catani:1993hr,Dokshitzer:1997in,Wobisch:1998wt,Cacciari:2008gp} just imposes an angular cut on the particles, and at this order, there are also no soft divergences.  Thus, we can calculate the cross section for this jet with mixed-quark flavor straightforwardly.  The cross section $\sigma^{(1)}_{pp\to W^+(bu)}(R,p_{\perp,W} > 1000\text{ GeV})$ as a function of radius $R$ is plotted in Fig.~\ref{fig:loflavxsec}.  Here, note that we use the superscript $(1)$ to denote that this is at next-to-leading order with respect to inclusive jet production.  However, this jet flavor configuration is calculated from tree-level diagrams.  As the matrix element is smooth in the limit that the $b$ and $u$ become collinear, the leading-order cross section for this flavor configuration decreases as the jet radius $R$ decreases.  The particular value of the cross section here isn't so important, but will be a benchmark for higher-order contributions to validate convergence of the perturbative expansion.

\subsubsection{Next-to-Leading Order Analysis}

At next-to-leading order, the cross section $\sigma^{(2)}_{pp\to W^+(bu)}(R,p_{\perp,W} > 1000\text{ GeV})$ of course receives contributions from virtual diagrams and real gluon emissions.  As long as the jet contains the $b$ and $u$ quarks, both of these contribute exclusively to the $(bu)$ flavor cross section.  Additionally, there are two other real emission contributions to this jet flavor channel: $pp\to W^+bu\bar u$, as already discussed, and $pp\to W^+bu\bar c$.  This latter process has no final state infrared or collinear divergences, and so this flavor channel's cross section can be evaluated from tree-level event generation, and then appropriate jet finding with {\sc FastJet} \cite{Cacciari:2011ma}, for example.

Unfortunately, accessing individual parton flavor information in inclusive event generation at next-to-leading order is not directly possible in many public, all-purpose programs like {\sc MadGraph} or {\sc MCFM} \cite{Campbell:2019dru}.\footnote{I thank Rikkert Frederix and Olivier Mattelaer for clarification of accessible information in {\sc MadGraph}.}  This will limit our ability to make general predictions at next-to-leading order for the cross section for generation of $(bu)$ flavored jets.  Nevertheless,  with our limited goal of validating our calculations of the flavor-dependent non-global logarithms, we can approach this problem in a roundabout way, without the need for generating exclusive events.  In {\sc MadGraph}, we can easily implement cuts on the minimum or maximum distance between the bottom quark and a light parton, as measured by the distance $\Delta R^2 = \Delta\eta^2+\Delta\phi^2$, where $\Delta\eta$ and $\Delta \phi$ are the differences in pseudorapidity and azimuthal angle, respectively.  We can exploit this feature and properties of the cross section in the soft limit to extract the desired information.

Specifically, this enables us to calculate the following three cross sections for the underlying process $pp\to W^+bu\bar u$:
\begin{enumerate}
\item the total cross section $\sigma(p_{\perp} > p_{\perp,\text{cut}})$ for $u,\bar u$ emission anywhere, with both of their transverse momenta to be larger than $p_{\perp,\text{cut}}$,
\item the cross section $\sigma(p_{\perp} > p_{\perp,\text{cut}},\Delta R > R)$ for both $u$ and $\bar u$ to lie at a distance $\Delta R > R$, with both of their transverse momenta to be larger than $p_{\perp,\text{cut}}$,
\item and the cross section $\sigma(p_{\perp} > p_{\perp,\text{cut}},\Delta R < R)$ for both $u$ and $\bar u$ to lie at a distance $\Delta R < R$, with both of their transverse momenta to be larger than $p_{\perp,\text{cut}}$.
\end{enumerate}
From these three cross sections, we can then evaluate the cross section for the $u$ quark to lie in the jet of radius $R$ and the $\bar u$ to lie outside, and vice-versa, where
\begin{align}\label{eq:subnglsmg}
&\sigma(p_{\perp} > p_{\perp,\text{cut}},\Delta R_{bu} < R<\Delta R_{b\bar u} ) + \sigma(p_{\perp} > p_{\perp,\text{cut}},\Delta R_{b\bar u} < R<\Delta R_{bu})\\
&\hspace{3cm} = \sigma(p_{\perp} > p_{\perp,\text{cut}}) - \sigma(p_{\perp} > p_{\perp,\text{cut}},\Delta R > R) - \sigma(p_{\perp} > p_{\perp,\text{cut}},\Delta R < R)\,.
\nonumber
\end{align}
In the soft limit, the matrix element is symmetric under $u\leftrightarrow \bar u$ exchange, and so the infrared divergences of the two cross sections $\sigma(p_{\perp} > p_{\perp,\text{cut}},\Delta R_{bu} < R<\Delta R_{b\bar u} )$ and $\sigma(p_{\perp} > p_{\perp,\text{cut}},\Delta R_{b\bar u} < R<\Delta R_{bu})$ are equal, as $p_{\perp,\text{cut}}\to 0$.  As such, we will be able to validate the non-global logarithms in this process.  However, the finite remainders may not necessarily be equal, and so this procedure is not sufficient to calculate the cross section of this contribution, according to the subtractive flavor algorithm.

Additionally, this angular cut to the bottom quark is not the anti-$k_T$ algorithm for arbitrary kinematics.  As discussed in the calculation of the flavor-dependent non-global logarithms, the anti-$k_T$ algorithm does reduce to such a simple angular cut on the soft particles about the hard particle.  However, this is another limitation of our analysis that prohibits a complete and self-consistent prediction of the subtractive flavor algorithm cross section.  We will work to address and solve this incompleteness in future work to make direct predictions with the subtractive flavor algorithm.

In {\sc MadGraph}, transverse momentum cuts on partons can only be applied universally.  This is in contrast to our approach to defining and calculating the non-global logarithms, where we only measure quantities on the jet of interest, and are ignorant as to anything outside the jet.  Specifically, in {\sc MadGraph}, we can impose the constraints $p_{\perp,1},p_{\perp,2}>p_{\perp,\text{cut}}$, while in our calculation we only imposed $p_{\perp,1}>p_{\perp,\text{cut}}$, where particle 1 is the soft quark that lies in the jet, and particle 2 is the soft quark outside the jet.  However, the difference between these two constraints is beyond the logarithmic accuracy we work.  In particular, note that the difference of the constraints is
\begin{align}
\Theta(p_{\perp,1}-p_{\perp,\text{cut}})-\Theta(p_{\perp,1}-p_{\perp,\text{cut}})\Theta(p_{\perp,2}-p_{\perp,\text{cut}})=\Theta(p_{\perp,1}-p_{\perp,\text{cut}})\Theta(p_{\perp,\text{cut}}-p_{\perp,2})\,.
\end{align}
With the variables we introduced in Sec.~\ref{sec:ngls}, where
\begin{align}
&p_{\perp,1}=p\,, &p_{\perp,2}=up\,,
\end{align}
we have to evaluate the integral over $p$ of the form
\begin{align}
\int \frac{dp}{p^{1+4\epsilon}}\,\Theta(p-p_{\perp,\text{cut}})\Theta(p_{\perp,\text{cut}}-up) = \log\frac{1}{u}+{\cal O}(\epsilon)\,.
\end{align}
The difference between the constraints we impose and what can be imposed in {\sc MadGraph} has no soft divergence, and is therefore subleading to the logarithmic accuracy we calculate.  Therefore, the {\sc MadGraph} constraints will result in the equivalent non-global logarithms, to the order we calculate.

This process has a single hard jet in the final state, so to calculate the non-global logarithms that contribute to the $(bu)$ flavor channel, we can use Eq.~\ref{eq:vjnglquad}.  For this final state, the Casimir of the $b$ jet is ${\bf T}_J^2 = C_F=4/3$ in QCD, while one beam must be a quark and the other a gluon.  Therefore, the sum of the Casimirs of the initial state is ${\bf T}_A^2+{\bf T}_B^2 = C_F+C_A$, where $C_A = 3$ in QCD.  With these color factors, the non-global logarithms are
\begin{align}\label{eq:nglpred}
\sigma^\text{(2)}_{u\bar u\text{,NGL}}(p_{\perp,\text{cut}})
&=\sigma_{pp\to W^+b}^{(0)}\left(
\frac{\alpha_s}{2\pi}
\right)^2T_R\,\log\frac{\mu}{p_{\perp,\text{cut}}}\left[
C_A\left(\frac{19}{9} R^2
-\frac{2}{3}R^2\log R^2\right)\right.\\
&\hspace{5.5cm}\left.+\,C_F\left(
-\frac{2}{3}+\frac{4}{9}\pi^2+\frac{R^2}{2} 
\right)
+{\cal O}(R^4)
\right]
\,.\nonumber
\end{align}
Not that, in particular, both terms proportional to $C_A$ and $C_F$ increase with increasing jet radius.  This expression still contains the renormalization scale $\mu$, which is also the scale at which the running coupling $\alpha_s$ is evaluated.  As the $W$ boson's transverse momentum is required to be larger than 1000 GeV, this is the order of the hard scale of the process, and therefore the natural choice for $\mu$.  Specifically, we take $\mu=1000$ GeV, and determine the value of $\alpha_s(\mu)$ at that scale through one-loop running.

\begin{figure}[t!]
\begin{center}
\includegraphics[width=0.45\textwidth]{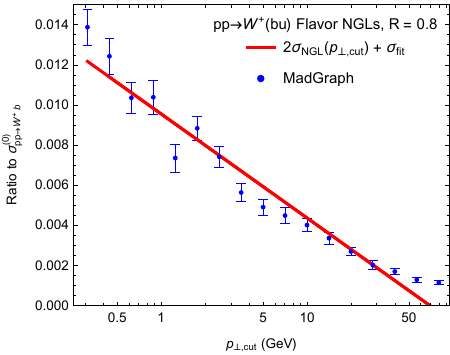}\hspace{1cm}
\includegraphics[width=0.45\textwidth]{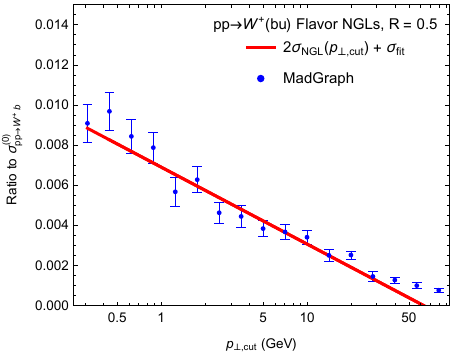}\\
\includegraphics[width=0.45\textwidth]{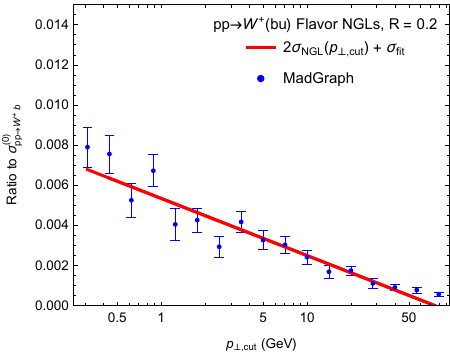}
\caption{\label{fig:nglmg}
Plots comparing the cross sections for non-global quark configurations in the process $pp\to W^+bu\bar u$, as a function of the transverse momentum cut $p_{\perp,\text{cut}}$ on the light quarks.  Three values of the jet radius about the direction of the $b$ quark are considered: $R=0.8$ (upper left), $R=0.5$ (upper right), and $R=0.2$ (bottom).  Cross sections calculated from {\sc MadGraph} sampling $10^4$ phase space points are plotted in blue with numerical uncertainty included, while the solid red line is two times the non-global logarithm prediction of Eq.~\ref{eq:nglpred}, plus a constant offset cross section.  All quantities are expressed as a fraction of the leading-order cross section of the process $pp\to W^+b$, Eq.~\ref{eq:lowb}.
}
\end{center}
\end{figure}

With these considerations, we can test this expression for the non-global logarithms to the output of {\sc MadGraph}, which is shown in Fig.~\ref{fig:nglmg}.  In these plots, results from {\sc MadGraph} are determined by the procedure of Eq.~\ref{eq:subnglsmg}, for three values of the jet radius $R = 0.8,0.5,0.2$.  The resulting cross section is then plotted for various values of $p_{\perp,\text{cut}}$ applied on both the $u$ and $\bar u$.  Additionally, we note that there is a collinear singularity between the $u$ and $\bar u$ if they are both in the jet or both are outside the jet.  Therefore, we must also impose a minimal angular cut between the $u$ and $\bar u$ in {\sc MadGraph}, which we set to be $\Delta R_{u\bar u} > 0.01$.  This is significantly smaller than the jet radii we consider and in the subtraction of Eq.~\ref{eq:subnglsmg}, the collinear divergence cancels anyway.  Finally, we also maintain the numerical uncertainties on the cross section to demonstrate some sense of the goodness-of-fit by eye, at least.\footnote{While with this approach we do not need to analyze exclusive events in {\sc MadGraph}, the numerical accuracy of the cross sections calculated by {\sc MadGraph} does depend on the requested number of events.  As this is a validation and not a precision study, our results use the default $10^4$ events for each cross section calculated in {\sc MadGraph}.}

On these plots, we also show our prediction of Eq.~\ref{eq:nglpred}.  Because of the inclusive cuts imposed in {\sc MadGraph}, either the $u$ or $\bar u$ is allowed to be the quark in the jet.  Therefore, we must multiply Eq.~\ref{eq:nglpred} by a factor of 2 to account for this inclusivity.  With default parton distributions, the value of the strong coupling at the $Z$ pole in {\sc MadGraph} is set to be $\alpha_s(m_Z) = 0.13$, which is what we use and then run to the scale $\mu=1000$ GeV.  Eq.~\ref{eq:nglpred} also only describes the logarithmic dependence on $p_{\perp,\text{cut}}$, and residual constant terms as $p_{\perp,\text{cut}}\to 0$ are not included.  As such, we add a constant off-set to the non-global logarithms that is determined by matching the {\sc MadGraph} output.  With these considerations, good agreement between the output of {\sc MadGraph} and our prediction for the non-global logarithms is observed.  In particular, the decrease in the slope as the jet radius decreases is matched well between our prediction and {\sc MadGraph}.  On these figures, the cross sections are plotted as a ratio to the leading order cross section for $pp\to W^+b$ production, Eq.~\ref{eq:lowb}.

Again, while the cuts we imposed in {\sc MadGraph} are insufficient to make a concrete prediction for the subtractive jet flavor for anti-$k_T$ jets at next-to-leading order, it is nevertheless interesting to observe how the procedure would work in the case at hand.  Specifically, with the predictions for the non-global logarithms, the contribution to the subtractive jet flavor cross section would be exactly that constant offset that we fit from {\sc MadGraph}.  To make the plots of Fig.~\ref{fig:nglmg}, the constant offset cross sections that remain after subtracting the non-global logarithms as $p_{\perp,\text{cut}}\to 0$ are:
\begin{align}
\sigma_\text{fit}^{(R = 0.8)}&=-6.84\times 10^{-6}\text{ fb}\,,\\
\sigma_\text{fit}^{(R = 0.5)}&=-5.24\times 10^{-6}\text{ fb}\,,\\
\sigma_\text{fit}^{(R = 0.2)}&=-3.65\times 10^{-6}\text{ fb}\,.
\end{align} 
Note that all of these residual cross sections are negative, suggesting that these non-global contributions work to {\it reduce} the non-partonic jet flavor assignments.  Again, as discussed earlier, to quantitatively show this requires honestly including the correct jet algorithm, disambiguating contributions from $u$ versus $\bar u$ flavor, and matching transverse momentum constraints.  We look forward to detailed predictions with the subtractive flavor algorithm in future work.

\section{Looking Towards an All-Orders Subtractive Flavor Algorithm}\label{sec:allords}

Our focus in this paper has been a fixed-order analysis of flavor-dependent non-global logarithms and subtractive jet flavor.  However, in a realistic application of a flavor algorithm, events and jets within them will consist of an arbitrary number of particles. As such, we need to describe how the subtractive jet flavor algorithm can be applied and used in these cases, especially looking forward to potential experimental applications of this procedure.  Our discussion here will be very limited, just to illustrate how this might work.\footnote{I thank Gavin Salam for suggesting this analysis.}

In particular, we will just consider the possible partonic configurations for a jet to be labeled as a gluon, according to subtractive jet flavor.  Specifically, this means that all quarks in the jet are either (1) sufficiently high energy and paired with their anti-quark, or (2) sufficiently low energy if not paired with their anti-quark inside the jet.  There is some subtlety with the issue of identical quark flavors, as we addressed in the previous section, but for this discussion we will ignore this.  Further, a quark--anti-quark pair in the jet contributes 0 net flavor anyway, so taking the $p_{\perp,\text{cut}} \to 0$ limit does not matter to these quarks.  So, to ensure that the jet is labeled as gluon flavor, we only need to ensure that all unpaired quarks are unresolved, with transverse momentum below $p_{\perp,\text{cut}}$.

With this observation, we can employ a few simplifying assumptions to actually directly calculate the cross section for all unpaired quarks in a jet to be unresolved.  First, we will work in the large number of quark flavors $n_f\to \infty$ limit, in which there are many individual flavors of quarks and so a jet never has more than 1 pair of a given flavor of quarks.  This is similar to simplifications that arise in the large number of colors limit, $N_c\to \infty$, when considering correlated gluon emissions.  Next, we will work to leading order in the small radius $R\to 0$ limit.  In this limit, we can ignore contributions to non-global logarithms that scale as positive powers of $R$.  This means that the corresponding non-global logarithms are universal, and only depend on the net color of the jet of interest.  With these approximations, the cross section for the non-global configuration when a quark is unresolved in the jet is
\begin{align}\label{eq:nglxsecso}
\sigma^\text{(2)}_{q\bar q\text{,NGL}}(p_{\perp,\text{cut}})
&=-2\,C_A\,n_f\,T_R\,\sigma_{pp\to g}^{(0)}\left(
\frac{\alpha_s}{2\pi}
\right)^2\,\log\frac{\mu}{p_{\perp,\text{cut}}}\left(
-\frac{2}{3}+\frac{4}{9}\pi^2
\right)
\,.
\end{align}
Here, $\sigma_{pp\to g}^{(0)}$ is the leading-order cross section for producing the gluon jet of interest, the factor of $2n_f$ accounts for any quark (or anti-quark) that could land in the jet, and the overall minus sign is because we require that the quark has sufficiently small transverse momentum, rather than sufficiently large transverse momentum.

Our assumptions actually enable us to very simply resum the contribution from an arbitrary number of unresolved quarks or anti-quarks.  With these approximations and in the limit in which all soft quark--anti-quark pairs are emitted off of the hard jet strongly-ordered in energy and collinear to the jet direction, we can easily write down a differential equation that describes the evolution of the non-global quark flavor, as a function of $p_{\perp,\text{cut}}$.  As $p_{\perp,\text{cut}}$ is decreased by a small amount $\delta p_\perp$, there are two possibilities: either the net quark flavor is unchanged or a new soft quark pair is emitted (or newly resolved) that is split across the boundary.  As gluons carry no flavor the net quark flavor is completely blind to arbitrary gluon emission, and so resolved gluons do not appear in the evolution equation.  Similarly, if both quarks in a soft $q\bar q$ pair stay in the jet or leave the jet, the net quark flavor is also unchanged.  That is, for both gluon emission and soft quark pairs that both land in (out of) the jet, there is exact real-virtual cancellation and no large logarithms are generated.

These observations motivate first the finite difference equation for the strongly-ordered gluon jet flavor cross section $\sigma^{\text{(so)}}_{g}(p_{\perp,\text{cut}})$, where
\begin{align}
\sigma^{\text{(so)}}_{g}(p_{\perp,\text{cut}}-\delta p_\perp) = \sigma^{\text{(so)}}_{g}(p_{\perp,\text{cut}})-\delta p_\perp\,\frac{2C_An_fT_R\left(
\frac{\alpha_s}{2\pi}
\right)^2\left(
-\frac{2}{3}+\frac{4}{9}\pi^2
\right)}{p_{\perp,\text{cut}}}\,\sigma^{\text{(so)}}_{g}(p_{\perp,\text{cut}})\,.
\end{align}
The first term on the right represents no new quark flavor resolved as $p_{\perp,\text{cut}}$ is decreased, while the second term describes the gluon flavor cross section lost by a newly resolved non-global $q\bar q$ pair in the interval $[p_{\perp,\text{cut}}-\delta p_\perp,p_{\perp,\text{cut}}]$.  This follows from differentiating Eq.~\ref{eq:nglxsecso}.  The solution of the resulting differential equation for gluon flavor with unresolved quarks split across the jet boundary is
\begin{align}
\sigma^{\text{(so)}}_{g}(p_{\perp,\text{cut}}) =\sigma_{pp\to g}^{(0)}\,\exp\left[
-2\,C_A\,n_f\,T_R\left(
\frac{\alpha_s}{2\pi}
\right)^2\log\frac{\mu}{p_{\perp,\text{cut}}}\left(
-\frac{2}{3}+\frac{4}{9}\pi^2
\right)
\right]\,,
\end{align}
where $\sigma_{pp\to g}^{(0)}$ is the cross section for gluon jet flavor at the high scale $\mu$.  Next, according to our prescription of subtractive jet flavor, we must subtract the non-global logarithm contribution to all orders.  With these approximations, the non-global contribution is
\begin{align}
\sigma^{\text{(so)}}_{\text{NGL},g}(p_{\perp,\text{cut}}) =\sigma_{pp\to g}^{(0)}\,\left\{\exp\left[
-2\,C_A\,n_f\,T_R\left(
\frac{\alpha_s}{2\pi}
\right)^2\,\log\frac{\mu}{p_{\perp,\text{cut}}}\left(
-\frac{2}{3}+\frac{4}{9}\pi^2
\right)
\right]-1\right\}\,.
\end{align}
Note that we must subtract off the first term of the exponential because that configuration has no emitted soft quarks and therefore also no non-global logarithms.

With these results, we then take the difference and the limit that $p_{\perp,\text{cut}}\to 0$.  The subtractive jet flavor in this limit to all orders simplifies dramatically:
\begin{align}
\sigma^{\text{(so)}}_{\text{SJF},g}(p_{\perp,\text{cut}}) =\lim_{p_{\perp,\text{cut}}\to 0}\left[\sigma^{\text{(so)}}_{g}(p_{\perp,\text{cut}})-\sigma^{\text{(so)}}_{\text{NGL},g}(p_{\perp,\text{cut}})\right] = \sigma_{pp\to g}^{(0)}\,.
\end{align}
The subtractive jet flavor prescription eliminates the contribution from non-global logarithms to all orders, reducing the gluon flavor cross section to its leading-order value.  Within these approximations, it is easy to see how higher fixed orders can be incorporated.  We just need to increase the accuracy of the overall cross section, and identify the non-global logarithms to all orders appropriately.  

From general arguments for the all-orders structure of cumulative cross sections of infrared and collinear safe observables \cite{Catani:1992ua}, we can write the general cross section for gluon flavor when all large logarithms of $p_{\perp,\text{cut}}$ are resummed as:
\begin{align}
\sigma_{g}(p_{\perp,\text{cut}}) = C(\alpha_s)\,R\left(\alpha_s,\alpha_s\log\frac{\mu}{p_{\perp,\text{cut}}}\right)+D(p_{\perp,\text{cut}})\,.
\end{align}
In this expression, all large logarithms of $p_{\perp,\text{cut}}$ are resummed in $R\left(\alpha_s,\alpha_s\log\frac{\mu}{p_{\perp,\text{cut}}}\right)$.  This also has the limiting behavior that
\begin{align}
\lim_{p_{\perp,\text{cut}}\to 0}R\left(\alpha_s,\alpha_s\log\frac{\mu}{p_{\perp,\text{cut}}}\right) = 0\,,
\end{align}
and it has the perturbative expansion
\begin{align}
R\left(\alpha_s,\alpha_s\log\frac{\mu}{p_{\perp,\text{cut}}}\right) = 1+{\cal O}\left(
\alpha_s^2\log\frac{\mu}{p_{\perp,\text{cut}}}
\right)\,.
\end{align}
The coefficient $C(\alpha_s)$ is the hard matching coefficient and encodes fixed-order corrections to the resummation, and has the perturbative expansion
\begin{align}
C(\alpha_s) = \sigma_{pp\to g}^{(0)} + {\cal O}(\alpha_s)\,.
\end{align}
Finally, the function $D(p_{\perp,\text{cut}})$ is regular as $p_{\perp,\text{cut}}\to 0$, and vanishes in that limit.  Note that all of these factors in this general cross section for the gluon flavor are well-defined and can be evaluated in (resummed) perturbation theory.

Now, according to subtractive jet flavor, we subtract those large logarithms from non-global soft quark emission.  This exclusively non-global contribution is just a part of the first term; namely,
\begin{align}
\sigma_{\text{NGL},g}(p_{\perp,\text{cut}}) = C(\alpha_s)\left[R\left(\alpha_s,\alpha_s\log\frac{\mu}{p_{\perp,\text{cut}}}\right)-1\right]\,.
\end{align}
For subtractive jet flavor, we then take the difference between the full cross section, and its non-global contribution, where
\begin{align}
\sigma_{\text{SJF},g}(p_{\perp,\text{cut}}) =\lim_{p_{\perp,\text{cut}}\to 0}\left[
\sigma_{g}(p_{\perp,\text{cut}})-\sigma_{\text{NGL},g}(p_{\perp,\text{cut}})
\right] = C(\alpha_s)\,.
\end{align}
That is, the subtractive gluon jet flavor is exactly just the hard matching coefficient for this jet production process.  As mentioned earlier, each of these components can be calculated in perturbation theory to whatever accuracy one is able.

\section{Conclusions}\label{sec:concs}

A na\"ive definition of jet flavor, the inclusive sum of quark flavors in a jet of interest, is not infrared and collinear safe, first becoming ill-defined at next-to-next-to-leading order.  The emissions responsible for this unsafety correspond to a soft $q\bar q$ pair that deposits a non-zero contribution to the jet flavor.  This emission configuration is exactly that which corresponds to non-global logarithms of an observable measured exclusively on the jet.  By appropriate isolation of the soft non-global contributions to the jet flavor, we can explicitly subtract the divergent logarithms, rendering the jet flavor well-defined and infrared safe.  These non-global logarithms can be calculated in full generality at next-to-next-to-leading order, as a series in powers of the jet radius $R$.  Only including terms in the expansion through order $R^2$ is accurate to within 5\% of the complete result for all $R \leq 1$, sufficient for many precision studies.  We call this procedure subtractive jet flavor, and using our non-global logarithm calculations, we demonstrated how to extract cross sections of jets with this flavor definition in fixed-order event generation codes.

In a realistic experimental analysis, of course partons are not observable; only hadrons are.  Further, hadron flavor is in general very challenging to identify at a modern collider, with bottom (or possibly charm) the exception.  Like other recent jet flavor definitions, subtractive jet flavor would be most practically useful for precision studies of heavy flavor, specifically for jets in which bottom hadrons are present.  Perturbatively, non-global soft quarks spoil the IRC safety of na\"ive jet flavor, but at the hadronic level, such soft quarks would be bound in color-singlet hadrons.  Further, unlike perturbation theory with massless quarks, hadrons all have finite masses and must have non-zero momentum to be detected.  Because of this, there are no non-global divergences of flavor at hadron level, and so there is nothing to subtract.\footnote{Another recent jet flavor proposal is similar in this way to subtractive jet flavor on hadronized events \cite{Generet:2025gdy}.} This suggests that the implementation of subtractive jet flavor on hadronized events is identical to the implementation of na\" ive jet flavor, to simply sum together the net flavors of the hadrons in the jet, as all observed hadrons would satisfy the $p_{\perp,\text{cut}}\to 0$ cut.  In some previous studies, na\"ive flavor jets have been shown to be very similar at the hadron level to fully IRC safe flavor definitions \cite{Caola:2023wpj,Behring:2025ilo}.  More study of fixed-order, resummed, and hadronized predictions of subtractive jet flavor are necessary, but this observation may provide justification for why na\"ive jet flavor seems to work reasonably well in practice.

The biggest limitation of our numerical study of subtractive jet flavor in {\sc MadGraph} was imposing appropriate phase space constraints on individual partons, in a flavor sensitive way.  For tree-level processes, this can be accomplished by event analysis and jet finding on unweighted events.  However, using public, open-source next-to-leading order event generators, this cannot obviously or simply be extended to analysis at higher orders.  Nevertheless, there are many constraints on the jet flavor cross sections, such as sum rules for total cross sections, and it is possible these can be exploited to completely constrain the flavored cross sections through next-to-leading order.  Non-partonic jet flavor, jets whose flavor is identified as a diquark for example, has infrared divergences starting at next-to-next-to-leading order from inclusive jet production, as studied here.  While we were able to extract subtractive jet flavor cross sections, uncertainties were still large, and significantly higher precision fixed-order calculations may be needed.

Non-global emission configurations and their contribution to the jet flavor can of course be identified order-by-order in perturbation theory.  However, if these flavor-dependent non-global logarithms could be systematically accounted for at fixed logarithmic accuracy but to all orders in the coupling through a factorization theorem, this would significantly simplify extraction of subtractive jet flavor cross sections.  It is possible that factorization and corresponding resummation is not too complicated, e.g., along the lines of Ref.~\cite{Becher:2023vrh} and related.  In particular, note that flavor-dependent non-global logarithms, where a pair of soft quarks are separated by a jet boundary, has a probability that scales like $\alpha_s^2 \log\frac{\mu}{p_{\perp,\text{cut}}}$.  As such, each time a soft quark pair is split across a jet boundary, the logarithmic accuracy of that contribution is pushed to a higher order.  Therefore, at fixed logarithmic accuracy, one only needs to consider a fixed number of split soft quark pairs, and then an arbitrary number of gluon emissions about them.  Of course, leading non-global logarithms in the leading color approximation are subtle and challenging enough to resum, but including soft quarks may be only a minor modification to that framework.

Along these lines, there are now many efforts to develop parton showers accurate to next-to-leading logarithm \cite{Hoche:2024dee,Preuss:2024vyu}, and even accurate to next-to-next-to-leading logarithmic order \cite{vanBeekveld:2023ivn}.  Non-global logarithms arising from soft gluon splitting that is separated by a phase space boundary arise first at next-to-leading logarithmic accuracy for a wide range of observables, and tests to validate that these non-global logarithms are correctly generated by these parton showers exist \cite{FerrarioRavasio:2023kyg}.  However, the accuracy of non-global logarithms within these general parton shower programs is limited to the leading color approximation, and so the flavor-dependent non-global logarithms we consider here are not guaranteed to be accurately described.  Even without an all-orders factorization theorem, our fixed-order results for jet radius dependence in quark--anti-quark non-global logarithms can provide an important and necessary accuracy benchmark for these programs.  We look forward to application of these results and a deeper understanding of flavor-dependent non-global logarithms.

\acknowledgements

I thank Thomas Becher for discussions of non-global logarithm factorization, Simone Marzani for many discussions about jet flavor, and Gavin Salam for comments and discussion of next-to-leading non-global logarithms in parton showers.

\appendix

\section{Collinear Fragmentation Functions}\label{app:fragfunc}

A finite energy cut on particles is not IRC safe, but this is a triviality that can be dealt with by absorbing collinear divergences into fragmentation functions and their renormalization factors.  We just illustrate how this is done here for one case, but this can be extended to all flavor configurations.  Consider a quark flavor jet in the collinear limit.  The quark and gluon that collinearly split are unresolved if they lie within an angle $R$, and are resolved if they are outside that.  If they are outside $R$, then the entire jet is tagged as quark flavor if its energy is sufficiently large.  With one-loop splitting functions, the corresponding calculation of the bare jet function is:
\begin{align}
J^{(1,\text{bare})}_q(\mu_0,R,p_{\perp,\text{cut}}) &= \frac{\alpha_sC_F}{2\pi}\left(
\frac{\mu_0^2}{p_\perp^2}
\right)^\epsilon\frac{1}{\Gamma(1-\epsilon)} \int \frac{d\theta^2}{(\theta^2)^{1+\epsilon}}\,dz\,z^{-2\epsilon}(1-z)^{-2\epsilon}\,  \left[
\frac{1+z^2}{1-z}-\epsilon(1-z)
\right]\\
&\hspace{5cm}
\times\left[
\Theta(R^2-\theta^2)+\Theta(\theta^2-R^2)\Theta(zp_\perp - p_{\perp,\text{cut}})
\right]\nonumber\\
&=\frac{\alpha_sC_F}{2\pi}\left(
\frac{\mu_0^2}{p_\perp^2R^2}
\right)^\epsilon \frac{1}{\epsilon}\left[
\frac{1}{2}\frac{p_{\perp,\text{cut}}}{p_{\perp}}\left(
2+\frac{p_{\perp,\text{cut}}}{p_{\perp}}
\right)+2\log\left(
1-\frac{p_{\perp,\text{cut}}}{p_{\perp}}
\right)+{\cal O}(\epsilon)
\right]\nonumber\,.
\end{align}
To renormalize this jet function through one-loop accuracy, we introduce the renormalization factor
\begin{align}
Z_q(\mu,\mu_0) = 1-\frac{\alpha_sC_F}{2\pi}\left(
\frac{\mu_0^2}{\mu^2}
\right)^\epsilon \frac{1}{\epsilon}\left[
\frac{1}{2}\frac{p_{\perp,\text{cut}}}{p_{\perp}}\left(
2+\frac{p_{\perp,\text{cut}}}{p_{\perp}}
\right)+2\log\left(
1-\frac{p_{\perp,\text{cut}}}{p_{\perp}}
\right)
\right]\,.
\end{align}
Then, the renormalized jet function through one-loop is
\begin{align}
J^{(\text{ren})}_q(\mu,R,p_{\perp,\text{cut}})&=Z_q(\mu,\mu_0)J^{(\text{bare})}_q(\mu_0,R,p_{\perp,\text{cut}})\\
&= 1+\frac{\alpha_sC_F}{2\pi}\left[
\frac{1}{2}\frac{p_{\perp,\text{cut}}}{p_{\perp}}\left(
2+\frac{p_{\perp,\text{cut}}}{p_{\perp}}
\right)+2\log\left(
1-\frac{p_{\perp,\text{cut}}}{p_{\perp}}
\right)
\right]\log
\frac{\mu^2}{p_\perp^2R^2}
+\cdots\nonumber\,,
\end{align}
where the ellipses ignore constants/terms non-logarithmic in the jet radius.  Importantly, note that logarithms of $R$ vanish as $p_{\perp,\text{cut}}\to 0$.  That is, subtractive jet flavor becomes IRC safe in that limit.

For completeness, we present the corresponding calculation for gluon flavor jets.  In this case, the bare jet function is
\begin{align}
J^{(1,\text{bare})}_g(\mu_0,R,p_{\perp,\text{cut}}) &= \frac{\alpha_sT_R}{2\pi}\left(
\frac{\mu_0^2}{p_\perp^2}
\right)^\epsilon\frac{1}{\Gamma(1-\epsilon)} \int \frac{d\theta^2}{(\theta^2)^{1+\epsilon}}\,dz\,z^{-2\epsilon}(1-z)^{-2\epsilon}\,  \left[
1-\frac{2z(1-z)}{1-\epsilon}
\right]\\
&\hspace{1cm}
\times\left[
\Theta(R^2-\theta^2)+\Theta(\theta^2-R^2)\Theta(zp_\perp - p_{\perp,\text{cut}})\Theta\left((1-z)p_\perp - p_{\perp,\text{cut}}\right)
\right]\nonumber\\
&=-\frac{\alpha_sT_R}{\pi}\left(
\frac{\mu_0^2}{p_\perp^2R^2}
\right)^\epsilon\frac{1}{\epsilon}  \left[
\frac{p_{\perp,\text{cut}}}{p_\perp}-\left(
\frac{p_{\perp,\text{cut}}}{p_\perp}
\right)^2+\frac{2}{3}\left(
\frac{p_{\perp,\text{cut}}}{p_\perp}
\right)^3+{\cal O}(\epsilon)
\right]\,.
\nonumber
\end{align}
Note that this too vanishes for $p_{\perp,\text{cut}}\to 0$, and agrees with the Winner Take All jet flavor \cite{Caletti:2022glq} results for $2p_{\perp,\text{cut}}=p_\perp$.

\bibliography{refs}

\begin{thebibliography}{65}%
\makeatletter
\providecommand \@ifxundefined [1]{%
 \@ifx{#1\undefined}
}%
\providecommand \@ifnum [1]{%
 \ifnum #1\expandafter \@firstoftwo
 \else \expandafter \@secondoftwo
 \fi
}%
\providecommand \@ifx [1]{%
 \ifx #1\expandafter \@firstoftwo
 \else \expandafter \@secondoftwo
 \fi
}%
\providecommand \natexlab [1]{#1}%
\providecommand \enquote  [1]{``#1''}%
\providecommand \bibnamefont  [1]{#1}%
\providecommand \bibfnamefont [1]{#1}%
\providecommand \citenamefont [1]{#1}%
\providecommand \href@noop [0]{\@secondoftwo}%
\providecommand \href [0]{\begingroup \@sanitize@url \@href}%
\providecommand \@href[1]{\@@startlink{#1}\@@href}%
\providecommand \@@href[1]{\endgroup#1\@@endlink}%
\providecommand \@sanitize@url [0]{\catcode `\\12\catcode `\$12\catcode
  `\&12\catcode `\#12\catcode `\^12\catcode `\_12\catcode `\%12\relax}%
\providecommand \@@startlink[1]{}%
\providecommand \@@endlink[0]{}%
\providecommand \url  [0]{\begingroup\@sanitize@url \@url }%
\providecommand \@url [1]{\endgroup\@href {#1}{\urlprefix }}%
\providecommand \urlprefix  [0]{URL }%
\providecommand \Eprint [0]{\href }%
\providecommand \doibase [0]{https://doi.org/}%
\providecommand \selectlanguage [0]{\@gobble}%
\providecommand \bibinfo  [0]{\@secondoftwo}%
\providecommand \bibfield  [0]{\@secondoftwo}%
\providecommand \translation [1]{[#1]}%
\providecommand \BibitemOpen [0]{}%
\providecommand \bibitemStop [0]{}%
\providecommand \bibitemNoStop [0]{.\EOS\space}%
\providecommand \EOS [0]{\spacefactor3000\relax}%
\providecommand \BibitemShut  [1]{\csname bibitem#1\endcsname}%
\let\auto@bib@innerbib\@empty
\bibitem [{\citenamefont {Salam}(2010)}]{Salam:2010nqg}%
  \BibitemOpen
  \bibfield  {author} {\bibinfo {author} {\bibfnamefont {G.~P.}\ \bibnamefont
  {Salam}},\ }\bibfield  {title} {\bibinfo {title} {{Towards Jetography}},\
  }\href {https://doi.org/10.1140/epjc/s10052-010-1314-6} {\bibfield  {journal}
  {\bibinfo  {journal} {Eur. Phys. J. C}\ }\textbf {\bibinfo {volume} {67}},\
  \bibinfo {pages} {637} (\bibinfo {year} {2010})},\ \Eprint
  {https://arxiv.org/abs/0906.1833} {arXiv:0906.1833 [hep-ph]} \BibitemShut
  {NoStop}%
\bibitem [{\citenamefont {Abdesselam}\ \emph {et~al.}(2011)\citenamefont
  {Abdesselam} \emph {et~al.}}]{Abdesselam:2010pt}%
  \BibitemOpen
  \bibfield  {author} {\bibinfo {author} {\bibfnamefont {A.}~\bibnamefont
  {Abdesselam}} \emph {et~al.},\ }\bibfield  {title} {\bibinfo {title}
  {{Boosted Objects: A Probe of Beyond the Standard Model Physics}},\ }\href
  {https://doi.org/10.1140/epjc/s10052-011-1661-y} {\bibfield  {journal}
  {\bibinfo  {journal} {Eur. Phys. J. C}\ }\textbf {\bibinfo {volume} {71}},\
  \bibinfo {pages} {1661} (\bibinfo {year} {2011})},\ \Eprint
  {https://arxiv.org/abs/1012.5412} {arXiv:1012.5412 [hep-ph]} \BibitemShut
  {NoStop}%
\bibitem [{\citenamefont {Altheimer}\ \emph {et~al.}(2012)\citenamefont
  {Altheimer} \emph {et~al.}}]{Altheimer:2012mn}%
  \BibitemOpen
  \bibfield  {author} {\bibinfo {author} {\bibfnamefont {A.}~\bibnamefont
  {Altheimer}} \emph {et~al.},\ }\bibfield  {title} {\bibinfo {title} {{Jet
  Substructure at the Tevatron and LHC: New results, new tools, new
  benchmarks}},\ }\href {https://doi.org/10.1088/0954-3899/39/6/063001}
  {\bibfield  {journal} {\bibinfo  {journal} {J. Phys. G}\ }\textbf {\bibinfo
  {volume} {39}},\ \bibinfo {pages} {063001} (\bibinfo {year} {2012})},\
  \Eprint {https://arxiv.org/abs/1201.0008} {arXiv:1201.0008 [hep-ph]}
  \BibitemShut {NoStop}%
\bibitem [{\citenamefont {Altheimer}\ \emph {et~al.}(2014)\citenamefont
  {Altheimer} \emph {et~al.}}]{Altheimer:2013yza}%
  \BibitemOpen
  \bibfield  {author} {\bibinfo {author} {\bibfnamefont {A.}~\bibnamefont
  {Altheimer}} \emph {et~al.},\ }\bibfield  {title} {\bibinfo {title} {{Boosted
  Objects and Jet Substructure at the LHC. Report of BOOST2012, held at IFIC
  Valencia, 23rd-27th of July 2012}},\ }\href
  {https://doi.org/10.1140/epjc/s10052-014-2792-8} {\bibfield  {journal}
  {\bibinfo  {journal} {Eur. Phys. J. C}\ }\textbf {\bibinfo {volume} {74}},\
  \bibinfo {pages} {2792} (\bibinfo {year} {2014})},\ \Eprint
  {https://arxiv.org/abs/1311.2708} {arXiv:1311.2708 [hep-ex]} \BibitemShut
  {NoStop}%
\bibitem [{\citenamefont {Larkoski}\ \emph {et~al.}(2020)\citenamefont
  {Larkoski}, \citenamefont {Moult},\ and\ \citenamefont
  {Nachman}}]{Larkoski:2017jix}%
  \BibitemOpen
  \bibfield  {author} {\bibinfo {author} {\bibfnamefont {A.~J.}\ \bibnamefont
  {Larkoski}}, \bibinfo {author} {\bibfnamefont {I.}~\bibnamefont {Moult}},\
  and\ \bibinfo {author} {\bibfnamefont {B.}~\bibnamefont {Nachman}},\
  }\bibfield  {title} {\bibinfo {title} {{Jet Substructure at the Large Hadron
  Collider: A Review of Recent Advances in Theory and Machine Learning}},\
  }\href {https://doi.org/10.1016/j.physrep.2019.11.001} {\bibfield  {journal}
  {\bibinfo  {journal} {Phys. Rept.}\ }\textbf {\bibinfo {volume} {841}},\
  \bibinfo {pages} {1} (\bibinfo {year} {2020})},\ \Eprint
  {https://arxiv.org/abs/1709.04464} {arXiv:1709.04464 [hep-ph]} \BibitemShut
  {NoStop}%
\bibitem [{\citenamefont {Kogler}\ \emph {et~al.}(2019)\citenamefont {Kogler}
  \emph {et~al.}}]{Kogler:2018hem}%
  \BibitemOpen
  \bibfield  {author} {\bibinfo {author} {\bibfnamefont {R.}~\bibnamefont
  {Kogler}} \emph {et~al.},\ }\bibfield  {title} {\bibinfo {title} {{Jet
  Substructure at the Large Hadron Collider: Experimental Review}},\ }\href
  {https://doi.org/10.1103/RevModPhys.91.045003} {\bibfield  {journal}
  {\bibinfo  {journal} {Rev. Mod. Phys.}\ }\textbf {\bibinfo {volume} {91}},\
  \bibinfo {pages} {045003} (\bibinfo {year} {2019})},\ \Eprint
  {https://arxiv.org/abs/1803.06991} {arXiv:1803.06991 [hep-ex]} \BibitemShut
  {NoStop}%
\bibitem [{\citenamefont {Marzani}\ \emph {et~al.}(2019)\citenamefont
  {Marzani}, \citenamefont {Soyez},\ and\ \citenamefont
  {Spannowsky}}]{Marzani:2019hun}%
  \BibitemOpen
  \bibfield  {author} {\bibinfo {author} {\bibfnamefont {S.}~\bibnamefont
  {Marzani}}, \bibinfo {author} {\bibfnamefont {G.}~\bibnamefont {Soyez}},\
  and\ \bibinfo {author} {\bibfnamefont {M.}~\bibnamefont {Spannowsky}},\
  }\href {https://doi.org/10.1007/978-3-030-15709-8} {\emph {\bibinfo {title}
  {{Looking inside jets: an introduction to jet substructure and boosted-object
  phenomenology}}}},\ Vol.\ \bibinfo {volume} {958}\ (\bibinfo  {publisher}
  {Springer},\ \bibinfo {year} {2019})\ \Eprint
  {https://arxiv.org/abs/1901.10342} {arXiv:1901.10342 [hep-ph]} \BibitemShut
  {NoStop}%
\bibitem [{\citenamefont {Dasgupta}\ and\ \citenamefont
  {Salam}(2001)}]{Dasgupta:2001sh}%
  \BibitemOpen
  \bibfield  {author} {\bibinfo {author} {\bibfnamefont {M.}~\bibnamefont
  {Dasgupta}}\ and\ \bibinfo {author} {\bibfnamefont {G.~P.}\ \bibnamefont
  {Salam}},\ }\bibfield  {title} {\bibinfo {title} {{Resummation of nonglobal
  QCD observables}},\ }\href {https://doi.org/10.1016/S0370-2693(01)00725-0}
  {\bibfield  {journal} {\bibinfo  {journal} {Phys. Lett. B}\ }\textbf
  {\bibinfo {volume} {512}},\ \bibinfo {pages} {323} (\bibinfo {year}
  {2001})},\ \Eprint {https://arxiv.org/abs/hep-ph/0104277}
  {arXiv:hep-ph/0104277} \BibitemShut {NoStop}%
\bibitem [{\citenamefont {Banfi}\ \emph {et~al.}(2002)\citenamefont {Banfi},
  \citenamefont {Marchesini},\ and\ \citenamefont {Smye}}]{Banfi:2002hw}%
  \BibitemOpen
  \bibfield  {author} {\bibinfo {author} {\bibfnamefont {A.}~\bibnamefont
  {Banfi}}, \bibinfo {author} {\bibfnamefont {G.}~\bibnamefont {Marchesini}},\
  and\ \bibinfo {author} {\bibfnamefont {G.}~\bibnamefont {Smye}},\ }\bibfield
  {title} {\bibinfo {title} {{Away from jet energy flow}},\ }\href
  {https://doi.org/10.1088/1126-6708/2002/08/006} {\bibfield  {journal}
  {\bibinfo  {journal} {JHEP}\ }\textbf {\bibinfo {volume} {08}},\ \bibinfo
  {pages} {006}},\ \Eprint {https://arxiv.org/abs/hep-ph/0206076}
  {arXiv:hep-ph/0206076} \BibitemShut {NoStop}%
\bibitem [{\citenamefont {Schwartz}\ and\ \citenamefont
  {Zhu}(2014)}]{Schwartz:2014wha}%
  \BibitemOpen
  \bibfield  {author} {\bibinfo {author} {\bibfnamefont {M.~D.}\ \bibnamefont
  {Schwartz}}\ and\ \bibinfo {author} {\bibfnamefont {H.~X.}\ \bibnamefont
  {Zhu}},\ }\bibfield  {title} {\bibinfo {title} {{Nonglobal logarithms at
  three loops, four loops, five loops, and beyond}},\ }\href
  {https://doi.org/10.1103/PhysRevD.90.065004} {\bibfield  {journal} {\bibinfo
  {journal} {Phys. Rev. D}\ }\textbf {\bibinfo {volume} {90}},\ \bibinfo
  {pages} {065004} (\bibinfo {year} {2014})},\ \Eprint
  {https://arxiv.org/abs/1403.4949} {arXiv:1403.4949 [hep-ph]} \BibitemShut
  {NoStop}%
\bibitem [{\citenamefont {Caron-Huot}(2018)}]{Caron-Huot:2015bja}%
  \BibitemOpen
  \bibfield  {author} {\bibinfo {author} {\bibfnamefont {S.}~\bibnamefont
  {Caron-Huot}},\ }\bibfield  {title} {\bibinfo {title} {{Resummation of
  non-global logarithms and the BFKL equation}},\ }\href
  {https://doi.org/10.1007/JHEP03(2018)036} {\bibfield  {journal} {\bibinfo
  {journal} {JHEP}\ }\textbf {\bibinfo {volume} {03}},\ \bibinfo {pages}
  {036}},\ \Eprint {https://arxiv.org/abs/1501.03754} {arXiv:1501.03754
  [hep-ph]} \BibitemShut {NoStop}%
\bibitem [{\citenamefont {Larkoski}\ \emph {et~al.}(2015)\citenamefont
  {Larkoski}, \citenamefont {Moult},\ and\ \citenamefont
  {Neill}}]{Larkoski:2015zka}%
  \BibitemOpen
  \bibfield  {author} {\bibinfo {author} {\bibfnamefont {A.~J.}\ \bibnamefont
  {Larkoski}}, \bibinfo {author} {\bibfnamefont {I.}~\bibnamefont {Moult}},\
  and\ \bibinfo {author} {\bibfnamefont {D.}~\bibnamefont {Neill}},\ }\bibfield
   {title} {\bibinfo {title} {{Non-Global Logarithms, Factorization, and the
  Soft Substructure of Jets}},\ }\href
  {https://doi.org/10.1007/JHEP09(2015)143} {\bibfield  {journal} {\bibinfo
  {journal} {JHEP}\ }\textbf {\bibinfo {volume} {09}},\ \bibinfo {pages}
  {143}},\ \Eprint {https://arxiv.org/abs/1501.04596} {arXiv:1501.04596
  [hep-ph]} \BibitemShut {NoStop}%
\bibitem [{\citenamefont {Becher}\ \emph {et~al.}(2016)\citenamefont {Becher},
  \citenamefont {Neubert}, \citenamefont {Rothen},\ and\ \citenamefont
  {Shao}}]{Becher:2015hka}%
  \BibitemOpen
  \bibfield  {author} {\bibinfo {author} {\bibfnamefont {T.}~\bibnamefont
  {Becher}}, \bibinfo {author} {\bibfnamefont {M.}~\bibnamefont {Neubert}},
  \bibinfo {author} {\bibfnamefont {L.}~\bibnamefont {Rothen}},\ and\ \bibinfo
  {author} {\bibfnamefont {D.~Y.}\ \bibnamefont {Shao}},\ }\bibfield  {title}
  {\bibinfo {title} {{Effective Field Theory for Jet Processes}},\ }\href
  {https://doi.org/10.1103/PhysRevLett.116.192001} {\bibfield  {journal}
  {\bibinfo  {journal} {Phys. Rev. Lett.}\ }\textbf {\bibinfo {volume} {116}},\
  \bibinfo {pages} {192001} (\bibinfo {year} {2016})},\ \Eprint
  {https://arxiv.org/abs/1508.06645} {arXiv:1508.06645 [hep-ph]} \BibitemShut
  {NoStop}%
\bibitem [{\citenamefont {\'Angeles~Mart\'\i{}nez}\ \emph
  {et~al.}(2018)\citenamefont {\'Angeles~Mart\'\i{}nez}, \citenamefont
  {De~Angelis}, \citenamefont {Forshaw}, \citenamefont {Pl\"atzer},\ and\
  \citenamefont {Seymour}}]{AngelesMartinez:2018cfz}%
  \BibitemOpen
  \bibfield  {author} {\bibinfo {author} {\bibfnamefont {R.}~\bibnamefont
  {\'Angeles~Mart\'\i{}nez}}, \bibinfo {author} {\bibfnamefont
  {M.}~\bibnamefont {De~Angelis}}, \bibinfo {author} {\bibfnamefont {J.~R.}\
  \bibnamefont {Forshaw}}, \bibinfo {author} {\bibfnamefont {S.}~\bibnamefont
  {Pl\"atzer}},\ and\ \bibinfo {author} {\bibfnamefont {M.~H.}\ \bibnamefont
  {Seymour}},\ }\bibfield  {title} {\bibinfo {title} {{Soft gluon evolution and
  non-global logarithms}},\ }\href {https://doi.org/10.1007/JHEP05(2018)044}
  {\bibfield  {journal} {\bibinfo  {journal} {JHEP}\ }\textbf {\bibinfo
  {volume} {05}},\ \bibinfo {pages} {044}},\ \Eprint
  {https://arxiv.org/abs/1802.08531} {arXiv:1802.08531 [hep-ph]} \BibitemShut
  {NoStop}%
\bibitem [{\citenamefont {Banfi}\ \emph {et~al.}(2022)\citenamefont {Banfi},
  \citenamefont {Dreyer},\ and\ \citenamefont {Monni}}]{Banfi:2021xzn}%
  \BibitemOpen
  \bibfield  {author} {\bibinfo {author} {\bibfnamefont {A.}~\bibnamefont
  {Banfi}}, \bibinfo {author} {\bibfnamefont {F.~A.}\ \bibnamefont {Dreyer}},\
  and\ \bibinfo {author} {\bibfnamefont {P.~F.}\ \bibnamefont {Monni}},\
  }\bibfield  {title} {\bibinfo {title} {{Higher-order non-global logarithms
  from jet calculus}},\ }\href {https://doi.org/10.1007/JHEP03(2022)135}
  {\bibfield  {journal} {\bibinfo  {journal} {JHEP}\ }\textbf {\bibinfo
  {volume} {03}},\ \bibinfo {pages} {135}},\ \Eprint
  {https://arxiv.org/abs/2111.02413} {arXiv:2111.02413 [hep-ph]} \BibitemShut
  {NoStop}%
\bibitem [{\citenamefont {Ferrario~Ravasio}\ \emph {et~al.}(2023)\citenamefont
  {Ferrario~Ravasio}, \citenamefont {Hamilton}, \citenamefont {Karlberg},
  \citenamefont {Salam}, \citenamefont {Scyboz},\ and\ \citenamefont
  {Soyez}}]{FerrarioRavasio:2023kyg}%
  \BibitemOpen
  \bibfield  {author} {\bibinfo {author} {\bibfnamefont {S.}~\bibnamefont
  {Ferrario~Ravasio}}, \bibinfo {author} {\bibfnamefont {K.}~\bibnamefont
  {Hamilton}}, \bibinfo {author} {\bibfnamefont {A.}~\bibnamefont {Karlberg}},
  \bibinfo {author} {\bibfnamefont {G.~P.}\ \bibnamefont {Salam}}, \bibinfo
  {author} {\bibfnamefont {L.}~\bibnamefont {Scyboz}},\ and\ \bibinfo {author}
  {\bibfnamefont {G.}~\bibnamefont {Soyez}},\ }\bibfield  {title} {\bibinfo
  {title} {{Parton Showering with Higher Logarithmic Accuracy for Soft
  Emissions}},\ }\href {https://doi.org/10.1103/PhysRevLett.131.161906}
  {\bibfield  {journal} {\bibinfo  {journal} {Phys. Rev. Lett.}\ }\textbf
  {\bibinfo {volume} {131}},\ \bibinfo {pages} {161906} (\bibinfo {year}
  {2023})},\ \Eprint {https://arxiv.org/abs/2307.11142} {arXiv:2307.11142
  [hep-ph]} \BibitemShut {NoStop}%
\bibitem [{\citenamefont {Becher}\ \emph {et~al.}(2024)\citenamefont {Becher},
  \citenamefont {Schalch},\ and\ \citenamefont {Xu}}]{Becher:2023vrh}%
  \BibitemOpen
  \bibfield  {author} {\bibinfo {author} {\bibfnamefont {T.}~\bibnamefont
  {Becher}}, \bibinfo {author} {\bibfnamefont {N.}~\bibnamefont {Schalch}},\
  and\ \bibinfo {author} {\bibfnamefont {X.}~\bibnamefont {Xu}},\ }\bibfield
  {title} {\bibinfo {title} {{Resummation of Next-to-Leading Nonglobal
  Logarithms at the LHC}},\ }\href
  {https://doi.org/10.1103/PhysRevLett.132.081602} {\bibfield  {journal}
  {\bibinfo  {journal} {Phys. Rev. Lett.}\ }\textbf {\bibinfo {volume} {132}},\
  \bibinfo {pages} {081602} (\bibinfo {year} {2024})},\ \Eprint
  {https://arxiv.org/abs/2307.02283} {arXiv:2307.02283 [hep-ph]} \BibitemShut
  {NoStop}%
\bibitem [{\citenamefont {Weigert}(2004)}]{Weigert:2003mm}%
  \BibitemOpen
  \bibfield  {author} {\bibinfo {author} {\bibfnamefont {H.}~\bibnamefont
  {Weigert}},\ }\bibfield  {title} {\bibinfo {title} {{Nonglobal jet evolution
  at finite N(c)}},\ }\href {https://doi.org/10.1016/j.nuclphysb.2004.03.002}
  {\bibfield  {journal} {\bibinfo  {journal} {Nucl. Phys. B}\ }\textbf
  {\bibinfo {volume} {685}},\ \bibinfo {pages} {321} (\bibinfo {year}
  {2004})},\ \Eprint {https://arxiv.org/abs/hep-ph/0312050}
  {arXiv:hep-ph/0312050} \BibitemShut {NoStop}%
\bibitem [{\citenamefont {Hatta}\ and\ \citenamefont
  {Ueda}(2013)}]{Hatta:2013iba}%
  \BibitemOpen
  \bibfield  {author} {\bibinfo {author} {\bibfnamefont {Y.}~\bibnamefont
  {Hatta}}\ and\ \bibinfo {author} {\bibfnamefont {T.}~\bibnamefont {Ueda}},\
  }\bibfield  {title} {\bibinfo {title} {{Resummation of non-global logarithms
  at finite $N_c$}},\ }\href {https://doi.org/10.1016/j.nuclphysb.2013.06.021}
  {\bibfield  {journal} {\bibinfo  {journal} {Nucl. Phys. B}\ }\textbf
  {\bibinfo {volume} {874}},\ \bibinfo {pages} {808} (\bibinfo {year}
  {2013})},\ \Eprint {https://arxiv.org/abs/1304.6930} {arXiv:1304.6930
  [hep-ph]} \BibitemShut {NoStop}%
\bibitem [{\citenamefont {Khelifa-Kerfa}\ and\ \citenamefont
  {Delenda}(2015)}]{Khelifa-Kerfa:2015mma}%
  \BibitemOpen
  \bibfield  {author} {\bibinfo {author} {\bibfnamefont {K.}~\bibnamefont
  {Khelifa-Kerfa}}\ and\ \bibinfo {author} {\bibfnamefont {Y.}~\bibnamefont
  {Delenda}},\ }\bibfield  {title} {\bibinfo {title} {{Non-global logarithms at
  finite N$_{c}$ beyond leading order}},\ }\href
  {https://doi.org/10.1007/JHEP03(2015)094} {\bibfield  {journal} {\bibinfo
  {journal} {JHEP}\ }\textbf {\bibinfo {volume} {03}},\ \bibinfo {pages}
  {094}},\ \Eprint {https://arxiv.org/abs/1501.00475} {arXiv:1501.00475
  [hep-ph]} \BibitemShut {NoStop}%
\bibitem [{\citenamefont {Kelley}\ \emph {et~al.}(2011)\citenamefont {Kelley},
  \citenamefont {Schwartz}, \citenamefont {Schabinger},\ and\ \citenamefont
  {Zhu}}]{Kelley:2011ng}%
  \BibitemOpen
  \bibfield  {author} {\bibinfo {author} {\bibfnamefont {R.}~\bibnamefont
  {Kelley}}, \bibinfo {author} {\bibfnamefont {M.~D.}\ \bibnamefont
  {Schwartz}}, \bibinfo {author} {\bibfnamefont {R.~M.}\ \bibnamefont
  {Schabinger}},\ and\ \bibinfo {author} {\bibfnamefont {H.~X.}\ \bibnamefont
  {Zhu}},\ }\bibfield  {title} {\bibinfo {title} {{The two-loop hemisphere soft
  function}},\ }\href {https://doi.org/10.1103/PhysRevD.84.045022} {\bibfield
  {journal} {\bibinfo  {journal} {Phys. Rev. D}\ }\textbf {\bibinfo {volume}
  {84}},\ \bibinfo {pages} {045022} (\bibinfo {year} {2011})},\ \Eprint
  {https://arxiv.org/abs/1105.3676} {arXiv:1105.3676 [hep-ph]} \BibitemShut
  {NoStop}%
\bibitem [{\citenamefont {Hornig}\ \emph {et~al.}(2011)\citenamefont {Hornig},
  \citenamefont {Lee}, \citenamefont {Stewart}, \citenamefont {Walsh},\ and\
  \citenamefont {Zuberi}}]{Hornig:2011iu}%
  \BibitemOpen
  \bibfield  {author} {\bibinfo {author} {\bibfnamefont {A.}~\bibnamefont
  {Hornig}}, \bibinfo {author} {\bibfnamefont {C.}~\bibnamefont {Lee}},
  \bibinfo {author} {\bibfnamefont {I.~W.}\ \bibnamefont {Stewart}}, \bibinfo
  {author} {\bibfnamefont {J.~R.}\ \bibnamefont {Walsh}},\ and\ \bibinfo
  {author} {\bibfnamefont {S.}~\bibnamefont {Zuberi}},\ }\bibfield  {title}
  {\bibinfo {title} {{Non-global Structure of the $\mathcal{O}({\alpha}^2_s)$
  Dijet Soft Function}},\ }\href {https://doi.org/10.1007/JHEP08(2011)054}
  {\bibfield  {journal} {\bibinfo  {journal} {JHEP}\ }\textbf {\bibinfo
  {volume} {08}},\ \bibinfo {pages} {054}},\ \bibinfo {note} {[Erratum: JHEP
  10, 101 (2017)]},\ \Eprint {https://arxiv.org/abs/1105.4628} {arXiv:1105.4628
  [hep-ph]} \BibitemShut {NoStop}%
\bibitem [{\citenamefont {Banfi}\ \emph {et~al.}(2006)\citenamefont {Banfi},
  \citenamefont {Salam},\ and\ \citenamefont {Zanderighi}}]{Banfi:2006hf}%
  \BibitemOpen
  \bibfield  {author} {\bibinfo {author} {\bibfnamefont {A.}~\bibnamefont
  {Banfi}}, \bibinfo {author} {\bibfnamefont {G.~P.}\ \bibnamefont {Salam}},\
  and\ \bibinfo {author} {\bibfnamefont {G.}~\bibnamefont {Zanderighi}},\
  }\bibfield  {title} {\bibinfo {title} {{Infrared safe definition of jet
  flavor}},\ }\href {https://doi.org/10.1140/epjc/s2006-02552-4} {\bibfield
  {journal} {\bibinfo  {journal} {Eur. Phys. J. C}\ }\textbf {\bibinfo {volume}
  {47}},\ \bibinfo {pages} {113} (\bibinfo {year} {2006})},\ \Eprint
  {https://arxiv.org/abs/hep-ph/0601139} {arXiv:hep-ph/0601139} \BibitemShut
  {NoStop}%
\bibitem [{\citenamefont {Caletti}\ \emph
  {et~al.}(2022{\natexlab{a}})\citenamefont {Caletti}, \citenamefont
  {Larkoski}, \citenamefont {Marzani},\ and\ \citenamefont
  {Reichelt}}]{Caletti:2022hnc}%
  \BibitemOpen
  \bibfield  {author} {\bibinfo {author} {\bibfnamefont {S.}~\bibnamefont
  {Caletti}}, \bibinfo {author} {\bibfnamefont {A.~J.}\ \bibnamefont
  {Larkoski}}, \bibinfo {author} {\bibfnamefont {S.}~\bibnamefont {Marzani}},\
  and\ \bibinfo {author} {\bibfnamefont {D.}~\bibnamefont {Reichelt}},\
  }\bibfield  {title} {\bibinfo {title} {{Practical jet flavour through
  NNLO}},\ }\href {https://doi.org/10.1140/epjc/s10052-022-10568-7} {\bibfield
  {journal} {\bibinfo  {journal} {Eur. Phys. J. C}\ }\textbf {\bibinfo {volume}
  {82}},\ \bibinfo {pages} {632} (\bibinfo {year} {2022}{\natexlab{a}})},\
  \Eprint {https://arxiv.org/abs/2205.01109} {arXiv:2205.01109 [hep-ph]}
  \BibitemShut {NoStop}%
\bibitem [{\citenamefont {Caletti}\ \emph
  {et~al.}(2022{\natexlab{b}})\citenamefont {Caletti}, \citenamefont
  {Larkoski}, \citenamefont {Marzani},\ and\ \citenamefont
  {Reichelt}}]{Caletti:2022glq}%
  \BibitemOpen
  \bibfield  {author} {\bibinfo {author} {\bibfnamefont {S.}~\bibnamefont
  {Caletti}}, \bibinfo {author} {\bibfnamefont {A.~J.}\ \bibnamefont
  {Larkoski}}, \bibinfo {author} {\bibfnamefont {S.}~\bibnamefont {Marzani}},\
  and\ \bibinfo {author} {\bibfnamefont {D.}~\bibnamefont {Reichelt}},\
  }\bibfield  {title} {\bibinfo {title} {{A fragmentation approach to jet
  flavor}},\ }\href {https://doi.org/10.1007/JHEP10(2022)158} {\bibfield
  {journal} {\bibinfo  {journal} {JHEP}\ }\textbf {\bibinfo {volume} {10}},\
  \bibinfo {pages} {158}},\ \Eprint {https://arxiv.org/abs/2205.01117}
  {arXiv:2205.01117 [hep-ph]} \BibitemShut {NoStop}%
\bibitem [{\citenamefont {Czakon}\ \emph {et~al.}(2023)\citenamefont {Czakon},
  \citenamefont {Mitov},\ and\ \citenamefont {Poncelet}}]{Czakon:2022wam}%
  \BibitemOpen
  \bibfield  {author} {\bibinfo {author} {\bibfnamefont {M.}~\bibnamefont
  {Czakon}}, \bibinfo {author} {\bibfnamefont {A.}~\bibnamefont {Mitov}},\ and\
  \bibinfo {author} {\bibfnamefont {R.}~\bibnamefont {Poncelet}},\ }\bibfield
  {title} {\bibinfo {title} {{Infrared-safe flavoured anti-k$_{T}$ jets}},\
  }\href {https://doi.org/10.1007/JHEP04(2023)138} {\bibfield  {journal}
  {\bibinfo  {journal} {JHEP}\ }\textbf {\bibinfo {volume} {04}},\ \bibinfo
  {pages} {138}},\ \Eprint {https://arxiv.org/abs/2205.11879} {arXiv:2205.11879
  [hep-ph]} \BibitemShut {NoStop}%
\bibitem [{\citenamefont {Gauld}\ \emph {et~al.}(2023)\citenamefont {Gauld},
  \citenamefont {Huss},\ and\ \citenamefont {Stagnitto}}]{Gauld:2022lem}%
  \BibitemOpen
  \bibfield  {author} {\bibinfo {author} {\bibfnamefont {R.}~\bibnamefont
  {Gauld}}, \bibinfo {author} {\bibfnamefont {A.}~\bibnamefont {Huss}},\ and\
  \bibinfo {author} {\bibfnamefont {G.}~\bibnamefont {Stagnitto}},\ }\bibfield
  {title} {\bibinfo {title} {{Flavor Identification of Reconstructed Hadronic
  Jets}},\ }\href {https://doi.org/10.1103/PhysRevLett.130.161901} {\bibfield
  {journal} {\bibinfo  {journal} {Phys. Rev. Lett.}\ }\textbf {\bibinfo
  {volume} {130}},\ \bibinfo {pages} {161901} (\bibinfo {year} {2023})},\
  \Eprint {https://arxiv.org/abs/2208.11138} {arXiv:2208.11138 [hep-ph]}
  \BibitemShut {NoStop}%
\bibitem [{\citenamefont {Caola}\ \emph {et~al.}(2023)\citenamefont {Caola},
  \citenamefont {Grabarczyk}, \citenamefont {Hutt}, \citenamefont {Salam},
  \citenamefont {Scyboz},\ and\ \citenamefont {Thaler}}]{Caola:2023wpj}%
  \BibitemOpen
  \bibfield  {author} {\bibinfo {author} {\bibfnamefont {F.}~\bibnamefont
  {Caola}}, \bibinfo {author} {\bibfnamefont {R.}~\bibnamefont {Grabarczyk}},
  \bibinfo {author} {\bibfnamefont {M.~L.}\ \bibnamefont {Hutt}}, \bibinfo
  {author} {\bibfnamefont {G.~P.}\ \bibnamefont {Salam}}, \bibinfo {author}
  {\bibfnamefont {L.}~\bibnamefont {Scyboz}},\ and\ \bibinfo {author}
  {\bibfnamefont {J.}~\bibnamefont {Thaler}},\ }\bibfield  {title} {\bibinfo
  {title} {{Flavored jets with exact anti-kt kinematics and tests of infrared
  and collinear safety}},\ }\href {https://doi.org/10.1103/PhysRevD.108.094010}
  {\bibfield  {journal} {\bibinfo  {journal} {Phys. Rev. D}\ }\textbf {\bibinfo
  {volume} {108}},\ \bibinfo {pages} {094010} (\bibinfo {year} {2023})},\
  \Eprint {https://arxiv.org/abs/2306.07314} {arXiv:2306.07314 [hep-ph]}
  \BibitemShut {NoStop}%
\bibitem [{\citenamefont {Behring}\ \emph {et~al.}(2025)\citenamefont {Behring}
  \emph {et~al.}}]{Behring:2025ilo}%
  \BibitemOpen
  \bibfield  {author} {\bibinfo {author} {\bibfnamefont {A.}~\bibnamefont
  {Behring}} \emph {et~al.},\ }\bibfield  {title} {\bibinfo {title} {{Flavoured
  jet algorithms: a comparative study}},\ }\href
  {https://doi.org/10.1007/JHEP09(2025)149} {\bibfield  {journal} {\bibinfo
  {journal} {JHEP}\ }\textbf {\bibinfo {volume} {09}},\ \bibinfo {pages}
  {149}},\ \Eprint {https://arxiv.org/abs/2506.13449} {arXiv:2506.13449
  [hep-ph]} \BibitemShut {NoStop}%
\bibitem [{\citenamefont {Aaij}\ \emph {et~al.}(2025)\citenamefont {Aaij} \emph
  {et~al.}}]{LHCb:2025tvf}%
  \BibitemOpen
  \bibfield  {author} {\bibinfo {author} {\bibfnamefont {R.}~\bibnamefont
  {Aaij}} \emph {et~al.} (\bibinfo {collaboration} {LHCb}),\ }\bibfield
  {title} {\bibinfo {title} {{First measurement of $b$-jet mass with and
  without grooming}},\ }\href@noop {} {\  (\bibinfo {year} {2025})},\ \Eprint
  {https://arxiv.org/abs/2505.11955} {arXiv:2505.11955 [hep-ex]} \BibitemShut
  {NoStop}%
\bibitem [{\citenamefont {Cacciari}\ \emph {et~al.}(2008)\citenamefont
  {Cacciari}, \citenamefont {Salam},\ and\ \citenamefont
  {Soyez}}]{Cacciari:2008gp}%
  \BibitemOpen
  \bibfield  {author} {\bibinfo {author} {\bibfnamefont {M.}~\bibnamefont
  {Cacciari}}, \bibinfo {author} {\bibfnamefont {G.~P.}\ \bibnamefont
  {Salam}},\ and\ \bibinfo {author} {\bibfnamefont {G.}~\bibnamefont {Soyez}},\
  }\bibfield  {title} {\bibinfo {title} {{The anti-$k_t$ jet clustering
  algorithm}},\ }\href {https://doi.org/10.1088/1126-6708/2008/04/063}
  {\bibfield  {journal} {\bibinfo  {journal} {JHEP}\ }\textbf {\bibinfo
  {volume} {04}},\ \bibinfo {pages} {063}},\ \Eprint
  {https://arxiv.org/abs/0802.1189} {arXiv:0802.1189 [hep-ph]} \BibitemShut
  {NoStop}%
\bibitem [{\citenamefont {Alwall}\ \emph {et~al.}(2014)\citenamefont {Alwall},
  \citenamefont {Frederix}, \citenamefont {Frixione}, \citenamefont {Hirschi},
  \citenamefont {Maltoni}, \citenamefont {Mattelaer}, \citenamefont {Shao},
  \citenamefont {Stelzer}, \citenamefont {Torrielli},\ and\ \citenamefont
  {Zaro}}]{Alwall:2014hca}%
  \BibitemOpen
  \bibfield  {author} {\bibinfo {author} {\bibfnamefont {J.}~\bibnamefont
  {Alwall}}, \bibinfo {author} {\bibfnamefont {R.}~\bibnamefont {Frederix}},
  \bibinfo {author} {\bibfnamefont {S.}~\bibnamefont {Frixione}}, \bibinfo
  {author} {\bibfnamefont {V.}~\bibnamefont {Hirschi}}, \bibinfo {author}
  {\bibfnamefont {F.}~\bibnamefont {Maltoni}}, \bibinfo {author} {\bibfnamefont
  {O.}~\bibnamefont {Mattelaer}}, \bibinfo {author} {\bibfnamefont {H.~S.}\
  \bibnamefont {Shao}}, \bibinfo {author} {\bibfnamefont {T.}~\bibnamefont
  {Stelzer}}, \bibinfo {author} {\bibfnamefont {P.}~\bibnamefont {Torrielli}},\
  and\ \bibinfo {author} {\bibfnamefont {M.}~\bibnamefont {Zaro}},\ }\bibfield
  {title} {\bibinfo {title} {{The automated computation of tree-level and
  next-to-leading order differential cross sections, and their matching to
  parton shower simulations}},\ }\href
  {https://doi.org/10.1007/JHEP07(2014)079} {\bibfield  {journal} {\bibinfo
  {journal} {JHEP}\ }\textbf {\bibinfo {volume} {07}},\ \bibinfo {pages}
  {079}},\ \Eprint {https://arxiv.org/abs/1405.0301} {arXiv:1405.0301 [hep-ph]}
  \BibitemShut {NoStop}%
\bibitem [{\citenamefont {Catani}\ and\ \citenamefont
  {Grazzini}(2000)}]{Catani:1999ss}%
  \BibitemOpen
  \bibfield  {author} {\bibinfo {author} {\bibfnamefont {S.}~\bibnamefont
  {Catani}}\ and\ \bibinfo {author} {\bibfnamefont {M.}~\bibnamefont
  {Grazzini}},\ }\bibfield  {title} {\bibinfo {title} {{Infrared factorization
  of tree level QCD amplitudes at the next-to-next-to-leading order and
  beyond}},\ }\href {https://doi.org/10.1016/S0550-3213(99)00778-6} {\bibfield
  {journal} {\bibinfo  {journal} {Nucl. Phys. B}\ }\textbf {\bibinfo {volume}
  {570}},\ \bibinfo {pages} {287} (\bibinfo {year} {2000})},\ \Eprint
  {https://arxiv.org/abs/hep-ph/9908523} {arXiv:hep-ph/9908523} \BibitemShut
  {NoStop}%
\bibitem [{\citenamefont {Aversa}\ \emph {et~al.}(1990)\citenamefont {Aversa},
  \citenamefont {Greco}, \citenamefont {Chiappetta},\ and\ \citenamefont
  {Guillet}}]{Aversa:1990uv}%
  \BibitemOpen
  \bibfield  {author} {\bibinfo {author} {\bibfnamefont {F.}~\bibnamefont
  {Aversa}}, \bibinfo {author} {\bibfnamefont {M.}~\bibnamefont {Greco}},
  \bibinfo {author} {\bibfnamefont {P.}~\bibnamefont {Chiappetta}},\ and\
  \bibinfo {author} {\bibfnamefont {J.~P.}\ \bibnamefont {Guillet}},\
  }\bibfield  {title} {\bibinfo {title} {{Jet inclusive production to O
  $(alpha-s^{3)}$ : Comparison with data}},\ }\href
  {https://doi.org/10.1103/PhysRevLett.65.401} {\bibfield  {journal} {\bibinfo
  {journal} {Phys. Rev. Lett.}\ }\textbf {\bibinfo {volume} {65}},\ \bibinfo
  {pages} {401} (\bibinfo {year} {1990})}\BibitemShut {NoStop}%
\bibitem [{\citenamefont {Jager}\ \emph {et~al.}(2004)\citenamefont {Jager},
  \citenamefont {Stratmann},\ and\ \citenamefont {Vogelsang}}]{Jager:2004jh}%
  \BibitemOpen
  \bibfield  {author} {\bibinfo {author} {\bibfnamefont {B.}~\bibnamefont
  {Jager}}, \bibinfo {author} {\bibfnamefont {M.}~\bibnamefont {Stratmann}},\
  and\ \bibinfo {author} {\bibfnamefont {W.}~\bibnamefont {Vogelsang}},\
  }\bibfield  {title} {\bibinfo {title} {{Single inclusive jet production in
  polarized $p p$ collisions at $O(alpha^3_s)$}},\ }\href
  {https://doi.org/10.1103/PhysRevD.70.034010} {\bibfield  {journal} {\bibinfo
  {journal} {Phys. Rev. D}\ }\textbf {\bibinfo {volume} {70}},\ \bibinfo
  {pages} {034010} (\bibinfo {year} {2004})},\ \Eprint
  {https://arxiv.org/abs/hep-ph/0404057} {arXiv:hep-ph/0404057} \BibitemShut
  {NoStop}%
\bibitem [{\citenamefont {Dasgupta}\ \emph {et~al.}(2008)\citenamefont
  {Dasgupta}, \citenamefont {Magnea},\ and\ \citenamefont
  {Salam}}]{Dasgupta:2007wa}%
  \BibitemOpen
  \bibfield  {author} {\bibinfo {author} {\bibfnamefont {M.}~\bibnamefont
  {Dasgupta}}, \bibinfo {author} {\bibfnamefont {L.}~\bibnamefont {Magnea}},\
  and\ \bibinfo {author} {\bibfnamefont {G.~P.}\ \bibnamefont {Salam}},\
  }\bibfield  {title} {\bibinfo {title} {{Non-perturbative QCD effects in jets
  at hadron colliders}},\ }\href
  {https://doi.org/10.1088/1126-6708/2008/02/055} {\bibfield  {journal}
  {\bibinfo  {journal} {JHEP}\ }\textbf {\bibinfo {volume} {02}},\ \bibinfo
  {pages} {055}},\ \Eprint {https://arxiv.org/abs/0712.3014} {arXiv:0712.3014
  [hep-ph]} \BibitemShut {NoStop}%
\bibitem [{\citenamefont {Hornig}\ \emph {et~al.}(2012)\citenamefont {Hornig},
  \citenamefont {Lee}, \citenamefont {Walsh},\ and\ \citenamefont
  {Zuberi}}]{Hornig:2011tg}%
  \BibitemOpen
  \bibfield  {author} {\bibinfo {author} {\bibfnamefont {A.}~\bibnamefont
  {Hornig}}, \bibinfo {author} {\bibfnamefont {C.}~\bibnamefont {Lee}},
  \bibinfo {author} {\bibfnamefont {J.~R.}\ \bibnamefont {Walsh}},\ and\
  \bibinfo {author} {\bibfnamefont {S.}~\bibnamefont {Zuberi}},\ }\bibfield
  {title} {\bibinfo {title} {{Double Non-Global Logarithms In-N-Out of Jets}},\
  }\href {https://doi.org/10.1007/JHEP01(2012)149} {\bibfield  {journal}
  {\bibinfo  {journal} {JHEP}\ }\textbf {\bibinfo {volume} {01}},\ \bibinfo
  {pages} {149}},\ \Eprint {https://arxiv.org/abs/1110.0004} {arXiv:1110.0004
  [hep-ph]} \BibitemShut {NoStop}%
\bibitem [{\citenamefont {Dasgupta}\ \emph {et~al.}(2012)\citenamefont
  {Dasgupta}, \citenamefont {Khelifa-Kerfa}, \citenamefont {Marzani},\ and\
  \citenamefont {Spannowsky}}]{Dasgupta:2012hg}%
  \BibitemOpen
  \bibfield  {author} {\bibinfo {author} {\bibfnamefont {M.}~\bibnamefont
  {Dasgupta}}, \bibinfo {author} {\bibfnamefont {K.}~\bibnamefont
  {Khelifa-Kerfa}}, \bibinfo {author} {\bibfnamefont {S.}~\bibnamefont
  {Marzani}},\ and\ \bibinfo {author} {\bibfnamefont {M.}~\bibnamefont
  {Spannowsky}},\ }\bibfield  {title} {\bibinfo {title} {{On jet mass
  distributions in Z+jet and dijet processes at the LHC}},\ }\href
  {https://doi.org/10.1007/JHEP10(2012)126} {\bibfield  {journal} {\bibinfo
  {journal} {JHEP}\ }\textbf {\bibinfo {volume} {10}},\ \bibinfo {pages}
  {126}},\ \Eprint {https://arxiv.org/abs/1207.1640} {arXiv:1207.1640 [hep-ph]}
  \BibitemShut {NoStop}%
\bibitem [{\citenamefont {Dasgupta}\ \emph {et~al.}(2013)\citenamefont
  {Dasgupta}, \citenamefont {Fregoso}, \citenamefont {Marzani},\ and\
  \citenamefont {Salam}}]{Dasgupta:2013ihk}%
  \BibitemOpen
  \bibfield  {author} {\bibinfo {author} {\bibfnamefont {M.}~\bibnamefont
  {Dasgupta}}, \bibinfo {author} {\bibfnamefont {A.}~\bibnamefont {Fregoso}},
  \bibinfo {author} {\bibfnamefont {S.}~\bibnamefont {Marzani}},\ and\ \bibinfo
  {author} {\bibfnamefont {G.~P.}\ \bibnamefont {Salam}},\ }\bibfield  {title}
  {\bibinfo {title} {{Towards an understanding of jet substructure}},\ }\href
  {https://doi.org/10.1007/JHEP09(2013)029} {\bibfield  {journal} {\bibinfo
  {journal} {JHEP}\ }\textbf {\bibinfo {volume} {09}},\ \bibinfo {pages}
  {029}},\ \Eprint {https://arxiv.org/abs/1307.0007} {arXiv:1307.0007 [hep-ph]}
  \BibitemShut {NoStop}%
\bibitem [{\citenamefont {Lepage}(1978)}]{Lepage:1977sw}%
  \BibitemOpen
  \bibfield  {author} {\bibinfo {author} {\bibfnamefont {G.~P.}\ \bibnamefont
  {Lepage}},\ }\bibfield  {title} {\bibinfo {title} {{A New Algorithm for
  Adaptive Multidimensional Integration}},\ }\href
  {https://doi.org/10.1016/0021-9991(78)90004-9} {\bibfield  {journal}
  {\bibinfo  {journal} {J. Comput. Phys.}\ }\textbf {\bibinfo {volume} {27}},\
  \bibinfo {pages} {192} (\bibinfo {year} {1978})}\BibitemShut {NoStop}%
\bibitem [{\citenamefont {Lepage}(1980)}]{Lepage:1980dq}%
  \BibitemOpen
  \bibfield  {author} {\bibinfo {author} {\bibfnamefont {G.~P.}\ \bibnamefont
  {Lepage}},\ }\bibfield  {title} {\bibinfo {title} {{VEGAS: AN ADAPTIVE
  MULTIDIMENSIONAL INTEGRATION PROGRAM}},\ }\href@noop {} {\  (\bibinfo {year}
  {1980})}\BibitemShut {NoStop}%
\bibitem [{\citenamefont {Hahn}(2005)}]{Hahn:2004fe}%
  \BibitemOpen
  \bibfield  {author} {\bibinfo {author} {\bibfnamefont {T.}~\bibnamefont
  {Hahn}},\ }\bibfield  {title} {\bibinfo {title} {{CUBA: A Library for
  multidimensional numerical integration}},\ }\href
  {https://doi.org/10.1016/j.cpc.2005.01.010} {\bibfield  {journal} {\bibinfo
  {journal} {Comput. Phys. Commun.}\ }\textbf {\bibinfo {volume} {168}},\
  \bibinfo {pages} {78} (\bibinfo {year} {2005})},\ \Eprint
  {https://arxiv.org/abs/hep-ph/0404043} {arXiv:hep-ph/0404043} \BibitemShut
  {NoStop}%
\bibitem [{\citenamefont {Hayrapetyan}\ \emph {et~al.}(2025)\citenamefont
  {Hayrapetyan} \emph {et~al.}}]{CMS:2025eyd}%
  \BibitemOpen
  \bibfield  {author} {\bibinfo {author} {\bibfnamefont {A.}~\bibnamefont
  {Hayrapetyan}} \emph {et~al.} (\bibinfo {collaboration} {CMS}),\ }\bibfield
  {title} {\bibinfo {title} {{A method for correcting the substructure of
  multiprong jets using the Lund jet plane}},\ }\href@noop {} {\  (\bibinfo
  {year} {2025})},\ \Eprint {https://arxiv.org/abs/2507.07775}
  {arXiv:2507.07775 [hep-ex]} \BibitemShut {NoStop}%
\bibitem [{\citenamefont {Chekhovsky}\ \emph {et~al.}(2025)\citenamefont
  {Chekhovsky} \emph {et~al.}}]{CMS:2025bxo}%
  \BibitemOpen
  \bibfield  {author} {\bibinfo {author} {\bibfnamefont {V.}~\bibnamefont
  {Chekhovsky}} \emph {et~al.} (\bibinfo {collaboration} {CMS}),\ }\bibfield
  {title} {\bibinfo {title} {{Search for top squarks in final states with many
  light-flavor jets and 0, 1, or 2 charged leptons in proton-proton collisions
  at $\sqrt{s}$ = 13 TeV}},\ }\href@noop {} {\  (\bibinfo {year} {2025})},\
  \Eprint {https://arxiv.org/abs/2506.08825} {arXiv:2506.08825 [hep-ex]}
  \BibitemShut {NoStop}%
\bibitem [{\citenamefont {Aad}\ \emph {et~al.}(2025{\natexlab{a}})\citenamefont
  {Aad} \emph {et~al.}}]{ATLAS:2025bpp}%
  \BibitemOpen
  \bibfield  {author} {\bibinfo {author} {\bibfnamefont {G.}~\bibnamefont
  {Aad}} \emph {et~al.} (\bibinfo {collaboration} {ATLAS}),\ }\bibfield
  {title} {\bibinfo {title} {{Measurement of the top quark mass with the ATLAS
  detector using $t\bar{t}$ events with a high transverse momentum top
  quark}},\ }\href {https://doi.org/10.1016/j.physletb.2025.139608} {\bibfield
  {journal} {\bibinfo  {journal} {Phys. Lett. B}\ }\textbf {\bibinfo {volume}
  {867}},\ \bibinfo {pages} {139608} (\bibinfo {year} {2025}{\natexlab{a}})},\
  \Eprint {https://arxiv.org/abs/2502.18216} {arXiv:2502.18216 [hep-ex]}
  \BibitemShut {NoStop}%
\bibitem [{\citenamefont {Aad}\ \emph {et~al.}(2025{\natexlab{b}})\citenamefont
  {Aad} \emph {et~al.}}]{ATLAS:2025rfm}%
  \BibitemOpen
  \bibfield  {author} {\bibinfo {author} {\bibfnamefont {G.}~\bibnamefont
  {Aad}} \emph {et~al.} (\bibinfo {collaboration} {ATLAS}),\ }\bibfield
  {title} {\bibinfo {title} {{Search for decays of the Higgs boson into scalar
  particles decaying into four or six $b$-quarks using $pp$ collisions at
  $\sqrt{s}= 13\,\mathrm{TeV}$ with the ATLAS detector}},\ }\href@noop {} {\
  (\bibinfo {year} {2025}{\natexlab{b}})},\ \Eprint
  {https://arxiv.org/abs/2507.01165} {arXiv:2507.01165 [hep-ex]} \BibitemShut
  {NoStop}%
\bibitem [{\citenamefont {Aad}\ \emph {et~al.}(2025{\natexlab{c}})\citenamefont
  {Aad} \emph {et~al.}}]{ATLAS:2025woc}%
  \BibitemOpen
  \bibfield  {author} {\bibinfo {author} {\bibfnamefont {G.}~\bibnamefont
  {Aad}} \emph {et~al.} (\bibinfo {collaboration} {ATLAS}),\ }\bibfield
  {title} {\bibinfo {title} {{Search for a new pseudoscalar decaying into a
  pair of bottom and antibottom quarks in top-associated production in
  $\sqrt{s}=13$~TeV proton{\textendash}proton collisions with the ATLAS
  detector}},\ }\href {https://doi.org/10.1140/epjc/s10052-025-14507-0}
  {\bibfield  {journal} {\bibinfo  {journal} {Eur. Phys. J. C}\ }\textbf
  {\bibinfo {volume} {85}},\ \bibinfo {pages} {886} (\bibinfo {year}
  {2025}{\natexlab{c}})},\ \Eprint {https://arxiv.org/abs/2503.17254}
  {arXiv:2503.17254 [hep-ex]} \BibitemShut {NoStop}%
\bibitem [{\citenamefont {Salam}(2024)}]{mod2flav}%
  \BibitemOpen
  \bibfield  {author} {\bibinfo {author} {\bibfnamefont {G.}~\bibnamefont
  {Salam}},\ }\href
  {https://conference.ippp.dur.ac.uk/event/1301/contributions/6818/attachments/5339/6936/Durham-flav-intro.pdf}
  {\bibinfo {title} {[jet] flavour and irc safety}} (\bibinfo {year}
  {2024})\BibitemShut {NoStop}%
\bibitem [{\citenamefont {Becher}\ and\ \citenamefont
  {Schwartz}(2008)}]{Becher:2008cf}%
  \BibitemOpen
  \bibfield  {author} {\bibinfo {author} {\bibfnamefont {T.}~\bibnamefont
  {Becher}}\ and\ \bibinfo {author} {\bibfnamefont {M.~D.}\ \bibnamefont
  {Schwartz}},\ }\bibfield  {title} {\bibinfo {title} {{A precise determination
  of $\alpha_s$ from LEP thrust data using effective field theory}},\ }\href
  {https://doi.org/10.1088/1126-6708/2008/07/034} {\bibfield  {journal}
  {\bibinfo  {journal} {JHEP}\ }\textbf {\bibinfo {volume} {07}},\ \bibinfo
  {pages} {034}},\ \Eprint {https://arxiv.org/abs/0803.0342} {arXiv:0803.0342
  [hep-ph]} \BibitemShut {NoStop}%
\bibitem [{\citenamefont {Gauld}\ \emph {et~al.}(2020)\citenamefont {Gauld},
  \citenamefont {Gehrmann-De~Ridder}, \citenamefont {Glover}, \citenamefont
  {Huss},\ and\ \citenamefont {Majer}}]{Gauld:2020deh}%
  \BibitemOpen
  \bibfield  {author} {\bibinfo {author} {\bibfnamefont {R.}~\bibnamefont
  {Gauld}}, \bibinfo {author} {\bibfnamefont {A.}~\bibnamefont
  {Gehrmann-De~Ridder}}, \bibinfo {author} {\bibfnamefont {E.~W.~N.}\
  \bibnamefont {Glover}}, \bibinfo {author} {\bibfnamefont {A.}~\bibnamefont
  {Huss}},\ and\ \bibinfo {author} {\bibfnamefont {I.}~\bibnamefont {Majer}},\
  }\bibfield  {title} {\bibinfo {title} {{Predictions for $Z$ -Boson Production
  in Association with a $b$-Jet at $\mathcal {O}(\alpha_s^3)$}},\ }\href
  {https://doi.org/10.1103/PhysRevLett.125.222002} {\bibfield  {journal}
  {\bibinfo  {journal} {Phys. Rev. Lett.}\ }\textbf {\bibinfo {volume} {125}},\
  \bibinfo {pages} {222002} (\bibinfo {year} {2020})},\ \Eprint
  {https://arxiv.org/abs/2005.03016} {arXiv:2005.03016 [hep-ph]} \BibitemShut
  {NoStop}%
\bibitem [{\citenamefont {Czakon}\ \emph {et~al.}(2021)\citenamefont {Czakon},
  \citenamefont {Mitov}, \citenamefont {Pellen},\ and\ \citenamefont
  {Poncelet}}]{Czakon:2020coa}%
  \BibitemOpen
  \bibfield  {author} {\bibinfo {author} {\bibfnamefont {M.}~\bibnamefont
  {Czakon}}, \bibinfo {author} {\bibfnamefont {A.}~\bibnamefont {Mitov}},
  \bibinfo {author} {\bibfnamefont {M.}~\bibnamefont {Pellen}},\ and\ \bibinfo
  {author} {\bibfnamefont {R.}~\bibnamefont {Poncelet}},\ }\bibfield  {title}
  {\bibinfo {title} {{NNLO QCD predictions for W+c-jet production at the
  LHC}},\ }\href {https://doi.org/10.1007/JHEP06(2021)100} {\bibfield
  {journal} {\bibinfo  {journal} {JHEP}\ }\textbf {\bibinfo {volume} {06}},\
  \bibinfo {pages} {100}},\ \Eprint {https://arxiv.org/abs/2011.01011}
  {arXiv:2011.01011 [hep-ph]} \BibitemShut {NoStop}%
\bibitem [{\citenamefont {Gehrmann-De~Ridder}\ \emph
  {et~al.}(2024)\citenamefont {Gehrmann-De~Ridder}, \citenamefont {Gehrmann},
  \citenamefont {Glover}, \citenamefont {Huss}, \citenamefont {Garcia},\ and\
  \citenamefont {Stagnitto}}]{Gehrmann-DeRidder:2023gdl}%
  \BibitemOpen
  \bibfield  {author} {\bibinfo {author} {\bibfnamefont {A.}~\bibnamefont
  {Gehrmann-De~Ridder}}, \bibinfo {author} {\bibfnamefont {T.}~\bibnamefont
  {Gehrmann}}, \bibinfo {author} {\bibfnamefont {E.~W.~N.}\ \bibnamefont
  {Glover}}, \bibinfo {author} {\bibfnamefont {A.}~\bibnamefont {Huss}},
  \bibinfo {author} {\bibfnamefont {A.~R.}\ \bibnamefont {Garcia}},\ and\
  \bibinfo {author} {\bibfnamefont {G.}~\bibnamefont {Stagnitto}},\ }\bibfield
  {title} {\bibinfo {title} {{Precise QCD predictions for W-boson production in
  association with a charm jet}},\ }\href
  {https://doi.org/10.1140/epjc/s10052-024-12715-8} {\bibfield  {journal}
  {\bibinfo  {journal} {Eur. Phys. J. C}\ }\textbf {\bibinfo {volume} {84}},\
  \bibinfo {pages} {361} (\bibinfo {year} {2024})},\ \Eprint
  {https://arxiv.org/abs/2311.14991} {arXiv:2311.14991 [hep-ph]} \BibitemShut
  {NoStop}%
\bibitem [{\citenamefont {Mazzitelli}\ \emph {et~al.}(2025)\citenamefont
  {Mazzitelli}, \citenamefont {Sotnikov},\ and\ \citenamefont
  {Wiesemann}}]{Mazzitelli:2024ura}%
  \BibitemOpen
  \bibfield  {author} {\bibinfo {author} {\bibfnamefont {J.}~\bibnamefont
  {Mazzitelli}}, \bibinfo {author} {\bibfnamefont {V.}~\bibnamefont
  {Sotnikov}},\ and\ \bibinfo {author} {\bibfnamefont {M.}~\bibnamefont
  {Wiesemann}},\ }\bibfield  {title} {\bibinfo {title}
  {{Next-to-next-to-leading order event generation for Z-boson production in
  association with a bottom-quark pair}},\ }\href
  {https://doi.org/10.1103/1kqt-s3vk} {\bibfield  {journal} {\bibinfo
  {journal} {Phys. Rev. D}\ }\textbf {\bibinfo {volume} {112}},\ \bibinfo
  {pages} {056031} (\bibinfo {year} {2025})},\ \Eprint
  {https://arxiv.org/abs/2404.08598} {arXiv:2404.08598 [hep-ph]} \BibitemShut
  {NoStop}%
\bibitem [{\citenamefont {Biello}\ \emph {et~al.}(2025)\citenamefont {Biello},
  \citenamefont {Mazzitelli}, \citenamefont {Sankar}, \citenamefont
  {Wiesemann},\ and\ \citenamefont {Zanderighi}}]{Biello:2024pgo}%
  \BibitemOpen
  \bibfield  {author} {\bibinfo {author} {\bibfnamefont {C.}~\bibnamefont
  {Biello}}, \bibinfo {author} {\bibfnamefont {J.}~\bibnamefont {Mazzitelli}},
  \bibinfo {author} {\bibfnamefont {A.}~\bibnamefont {Sankar}}, \bibinfo
  {author} {\bibfnamefont {M.}~\bibnamefont {Wiesemann}},\ and\ \bibinfo
  {author} {\bibfnamefont {G.}~\bibnamefont {Zanderighi}},\ }\bibfield  {title}
  {\bibinfo {title} {{Higgs boson production in association with massive bottom
  quarks at NNLO+PS}},\ }\href {https://doi.org/10.1007/JHEP04(2025)088}
  {\bibfield  {journal} {\bibinfo  {journal} {JHEP}\ }\textbf {\bibinfo
  {volume} {04}},\ \bibinfo {pages} {088}},\ \Eprint
  {https://arxiv.org/abs/2412.09510} {arXiv:2412.09510 [hep-ph]} \BibitemShut
  {NoStop}%
\bibitem [{\citenamefont {Ellis}\ and\ \citenamefont
  {Soper}(1993)}]{Ellis:1993tq}%
  \BibitemOpen
  \bibfield  {author} {\bibinfo {author} {\bibfnamefont {S.~D.}\ \bibnamefont
  {Ellis}}\ and\ \bibinfo {author} {\bibfnamefont {D.~E.}\ \bibnamefont
  {Soper}},\ }\bibfield  {title} {\bibinfo {title} {{Successive combination jet
  algorithm for hadron collisions}},\ }\href
  {https://doi.org/10.1103/PhysRevD.48.3160} {\bibfield  {journal} {\bibinfo
  {journal} {Phys. Rev. D}\ }\textbf {\bibinfo {volume} {48}},\ \bibinfo
  {pages} {3160} (\bibinfo {year} {1993})},\ \Eprint
  {https://arxiv.org/abs/hep-ph/9305266} {arXiv:hep-ph/9305266} \BibitemShut
  {NoStop}%
\bibitem [{\citenamefont {Catani}\ \emph
  {et~al.}(1993{\natexlab{a}})\citenamefont {Catani}, \citenamefont
  {Dokshitzer}, \citenamefont {Seymour},\ and\ \citenamefont
  {Webber}}]{Catani:1993hr}%
  \BibitemOpen
  \bibfield  {author} {\bibinfo {author} {\bibfnamefont {S.}~\bibnamefont
  {Catani}}, \bibinfo {author} {\bibfnamefont {Y.~L.}\ \bibnamefont
  {Dokshitzer}}, \bibinfo {author} {\bibfnamefont {M.~H.}\ \bibnamefont
  {Seymour}},\ and\ \bibinfo {author} {\bibfnamefont {B.~R.}\ \bibnamefont
  {Webber}},\ }\bibfield  {title} {\bibinfo {title} {{Longitudinally invariant
  $K_t$ clustering algorithms for hadron hadron collisions}},\ }\href
  {https://doi.org/10.1016/0550-3213(93)90166-M} {\bibfield  {journal}
  {\bibinfo  {journal} {Nucl. Phys. B}\ }\textbf {\bibinfo {volume} {406}},\
  \bibinfo {pages} {187} (\bibinfo {year} {1993}{\natexlab{a}})}\BibitemShut
  {NoStop}%
\bibitem [{\citenamefont {Dokshitzer}\ \emph {et~al.}(1997)\citenamefont
  {Dokshitzer}, \citenamefont {Leder}, \citenamefont {Moretti},\ and\
  \citenamefont {Webber}}]{Dokshitzer:1997in}%
  \BibitemOpen
  \bibfield  {author} {\bibinfo {author} {\bibfnamefont {Y.~L.}\ \bibnamefont
  {Dokshitzer}}, \bibinfo {author} {\bibfnamefont {G.~D.}\ \bibnamefont
  {Leder}}, \bibinfo {author} {\bibfnamefont {S.}~\bibnamefont {Moretti}},\
  and\ \bibinfo {author} {\bibfnamefont {B.~R.}\ \bibnamefont {Webber}},\
  }\bibfield  {title} {\bibinfo {title} {{Better jet clustering algorithms}},\
  }\href {https://doi.org/10.1088/1126-6708/1997/08/001} {\bibfield  {journal}
  {\bibinfo  {journal} {JHEP}\ }\textbf {\bibinfo {volume} {08}},\ \bibinfo
  {pages} {001}},\ \Eprint {https://arxiv.org/abs/hep-ph/9707323}
  {arXiv:hep-ph/9707323} \BibitemShut {NoStop}%
\bibitem [{\citenamefont {Wobisch}\ and\ \citenamefont
  {Wengler}(1998)}]{Wobisch:1998wt}%
  \BibitemOpen
  \bibfield  {author} {\bibinfo {author} {\bibfnamefont {M.}~\bibnamefont
  {Wobisch}}\ and\ \bibinfo {author} {\bibfnamefont {T.}~\bibnamefont
  {Wengler}},\ }\bibfield  {title} {\bibinfo {title} {{Hadronization
  corrections to jet cross-sections in deep inelastic scattering}},\ }in\
  \href@noop {} {\emph {\bibinfo {booktitle} {{Workshop on Monte Carlo
  Generators for HERA Physics (Plenary Starting Meeting)}}}}\ (\bibinfo {year}
  {1998})\ pp.\ \bibinfo {pages} {270--279},\ \Eprint
  {https://arxiv.org/abs/hep-ph/9907280} {arXiv:hep-ph/9907280} \BibitemShut
  {NoStop}%
\bibitem [{\citenamefont {Cacciari}\ \emph {et~al.}(2012)\citenamefont
  {Cacciari}, \citenamefont {Salam},\ and\ \citenamefont
  {Soyez}}]{Cacciari:2011ma}%
  \BibitemOpen
  \bibfield  {author} {\bibinfo {author} {\bibfnamefont {M.}~\bibnamefont
  {Cacciari}}, \bibinfo {author} {\bibfnamefont {G.~P.}\ \bibnamefont
  {Salam}},\ and\ \bibinfo {author} {\bibfnamefont {G.}~\bibnamefont {Soyez}},\
  }\bibfield  {title} {\bibinfo {title} {{FastJet User Manual}},\ }\href
  {https://doi.org/10.1140/epjc/s10052-012-1896-2} {\bibfield  {journal}
  {\bibinfo  {journal} {Eur. Phys. J. C}\ }\textbf {\bibinfo {volume} {72}},\
  \bibinfo {pages} {1896} (\bibinfo {year} {2012})},\ \Eprint
  {https://arxiv.org/abs/1111.6097} {arXiv:1111.6097 [hep-ph]} \BibitemShut
  {NoStop}%
\bibitem [{\citenamefont {Campbell}\ and\ \citenamefont
  {Neumann}(2019)}]{Campbell:2019dru}%
  \BibitemOpen
  \bibfield  {author} {\bibinfo {author} {\bibfnamefont {J.}~\bibnamefont
  {Campbell}}\ and\ \bibinfo {author} {\bibfnamefont {T.}~\bibnamefont
  {Neumann}},\ }\bibfield  {title} {\bibinfo {title} {{Precision Phenomenology
  with MCFM}},\ }\href {https://doi.org/10.1007/JHEP12(2019)034} {\bibfield
  {journal} {\bibinfo  {journal} {JHEP}\ }\textbf {\bibinfo {volume} {12}},\
  \bibinfo {pages} {034}},\ \Eprint {https://arxiv.org/abs/1909.09117}
  {arXiv:1909.09117 [hep-ph]} \BibitemShut {NoStop}%
\bibitem [{\citenamefont {Catani}\ \emph
  {et~al.}(1993{\natexlab{b}})\citenamefont {Catani}, \citenamefont
  {Trentadue}, \citenamefont {Turnock},\ and\ \citenamefont
  {Webber}}]{Catani:1992ua}%
  \BibitemOpen
  \bibfield  {author} {\bibinfo {author} {\bibfnamefont {S.}~\bibnamefont
  {Catani}}, \bibinfo {author} {\bibfnamefont {L.}~\bibnamefont {Trentadue}},
  \bibinfo {author} {\bibfnamefont {G.}~\bibnamefont {Turnock}},\ and\ \bibinfo
  {author} {\bibfnamefont {B.~R.}\ \bibnamefont {Webber}},\ }\bibfield  {title}
  {\bibinfo {title} {{Resummation of large logarithms in e+ e- event shape
  distributions}},\ }\href {https://doi.org/10.1016/0550-3213(93)90271-P}
  {\bibfield  {journal} {\bibinfo  {journal} {Nucl. Phys. B}\ }\textbf
  {\bibinfo {volume} {407}},\ \bibinfo {pages} {3} (\bibinfo {year}
  {1993}{\natexlab{b}})}\BibitemShut {NoStop}%
\bibitem [{\citenamefont {Generet}(2025)}]{Generet:2025gdy}%
  \BibitemOpen
  \bibfield  {author} {\bibinfo {author} {\bibfnamefont {T.}~\bibnamefont
  {Generet}},\ }\bibfield  {title} {\bibinfo {title} {{IRC-safe jet flavour
  without modifying anything}},\ }\href@noop {} {\  (\bibinfo {year} {2025})},\
  \Eprint {https://arxiv.org/abs/2511.23423} {arXiv:2511.23423 [hep-ph]}
  \BibitemShut {NoStop}%
\bibitem [{\citenamefont {H{\"o}che}\ \emph {et~al.}(2025)\citenamefont
  {H{\"o}che}, \citenamefont {Krauss},\ and\ \citenamefont
  {Reichelt}}]{Hoche:2024dee}%
  \BibitemOpen
  \bibfield  {author} {\bibinfo {author} {\bibfnamefont {S.}~\bibnamefont
  {H{\"o}che}}, \bibinfo {author} {\bibfnamefont {F.}~\bibnamefont {Krauss}},\
  and\ \bibinfo {author} {\bibfnamefont {D.}~\bibnamefont {Reichelt}},\
  }\bibfield  {title} {\bibinfo {title} {{alaric parton shower for hadron
  colliders}},\ }\href {https://doi.org/10.1103/PhysRevD.111.094032} {\bibfield
   {journal} {\bibinfo  {journal} {Phys. Rev. D}\ }\textbf {\bibinfo {volume}
  {111}},\ \bibinfo {pages} {094032} (\bibinfo {year} {2025})},\ \Eprint
  {https://arxiv.org/abs/2404.14360} {arXiv:2404.14360 [hep-ph]} \BibitemShut
  {NoStop}%
\bibitem [{\citenamefont {Preuss}(2024)}]{Preuss:2024vyu}%
  \BibitemOpen
  \bibfield  {author} {\bibinfo {author} {\bibfnamefont {C.~T.}\ \bibnamefont
  {Preuss}},\ }\bibfield  {title} {\bibinfo {title} {{A partitioned
  dipole-antenna shower with improved transverse recoil}},\ }\href@noop {} {\
  (\bibinfo {year} {2024})},\ \Eprint {https://arxiv.org/abs/2403.19452}
  {arXiv:2403.19452 [hep-ph]} \BibitemShut {NoStop}%
\bibitem [{\citenamefont {van Beekveld}\ \emph {et~al.}(2024)\citenamefont {van
  Beekveld} \emph {et~al.}}]{vanBeekveld:2023ivn}%
  \BibitemOpen
  \bibfield  {author} {\bibinfo {author} {\bibfnamefont {M.}~\bibnamefont {van
  Beekveld}} \emph {et~al.},\ }\bibfield  {title} {\bibinfo {title}
  {{Introduction to the PanScales framework, version 0.1}},\ }\href
  {https://doi.org/10.21468/SciPostPhysCodeb.31} {\bibfield  {journal}
  {\bibinfo  {journal} {SciPost Phys. Codeb.}\ }\textbf {\bibinfo {volume}
  {2024}},\ \bibinfo {pages} {31} (\bibinfo {year} {2024})},\ \Eprint
  {https://arxiv.org/abs/2312.13275} {arXiv:2312.13275 [hep-ph]} \BibitemShut
  {NoStop}%
\end{thebibliography}%

\end{document}